\documentclass[twocolumn,aps,prd]{revtex4}
\usepackage{graphics,amsmath} 
\usepackage[dvips]{color}
\newcommand {\boldtheta}{\mbox{\boldmath $\theta$}}
\newcommand {\mpl}{M_\mathrm{Pl}}

\begin{document} 

\title{Cosmological Constraints on Dissipative Models of Inflation} 
\author{Lisa M. H. Hall}
\email{lisa.hall@sheffield.ac.uk}
\affiliation{Department of Applied Mathematics,
The University of Sheffield,
Hounsfield Road, Sheffield S3 7RH, UK}
\author{Hiranya V. Peiris\footnote{Hubble Fellow}}
\email{hiranya@ast.cam.ac.uk}
\affiliation{Kavli Institute for Cosmological Physics and Enrico Fermi Institute, University of Chicago, Chicago IL 60637, USA\\ and \\ Institute of Astronomy, University of Cambridge, Cambridge CB3 0HA, UK}

\date{\today}


\begin{abstract}
We study dissipative inflation in the regime where the dissipative term takes a specific form, $\Gamma=\Gamma(\phi)$, analyzing two models in the weak and strong dissipative regimes with a SUSY breaking potential. This system introduces three new parameters; two for the potential and one for the dissipative term. After developing intuition about the predictions from these models through analytic approximations, we compute the predicted cosmological observables through full numerical evolution of the equations of motion, relating the mass scale and scale of dissipation to the characteristic amplitude and shape of the primordial power spectrum. We then use Markov Chain Monte Carlo techniques to constrain a subset of the models with cosmological data from the cosmic microwave background (WMAP three-year data) and large scale structure (SDSS Luminous Red Galaxy power spectrum). We find that the primordial dissipative parameters are uncorrelated with the ``late-time'' cosmological parameters which describe the contents and expansion rate of the universe; the latter show no significant shift from the standard $\Lambda$CDM concordance cosmology and possess close to Gaussian posterior probability distributions. In contrast, the posterior distributions of the dissipative parameters are highly non-Gaussian and their allowed ranges agree well with the expectations obtained using analytic approximations.
In the weak regime, only the mass scale is tightly constrained; conversely, in the strong regime, only the dissipative coefficient is tightly constrained. A lower limit is seen on the inflation scale: a sub-Planckian inflaton is disfavoured by the data. In both weak and strong regimes, we reconstruct the limits on the primordial power spectrum and show that these models prefer a {\it red} spectrum, with no significant running of the index. Despite having one extra degree of freedom in the fit compared to the standard $\Lambda$CDM model, the data does not display a preference for any of the dissipative models; the goodness-of-fit is comparable to the latter. We calculate the reheat temperature and show that the gravitino problem can be overcome with large dissipation, which in turn leads to large levels of non-Gaussianity:  if dissipative inflation is to evade the gravitino problem, the predicted level of non-Gaussianity might be seen by the Planck satellite.
\end{abstract}
\pacs{PACS number(s): }

\maketitle
\section{introduction}

In the original model of inflation, the early universe is hypothesized to undergo
exponential expansion and supercool as a result of the radiation redshifting \cite{Guth:1981zm}. A subsequent, independent phase of reheating is required to return the universe
to a radiation dominated state \cite{Albrecht:1982mp,Dolgov:1982th,Abbott:1982hn,Kofman:1994rk,Kofman:1997yn}.

This original picture assumes that any interaction of the inflationary scalar field (called the \emph{inflaton}) has a negligible consequence on inflation. It has, however, been shown that a scalar field interacting with a thermal bath leads to an additional friction term in the equation of motion \cite{Hosoya:1984ke}. Hence, if the universe begins in a thermally excited state, the inflaton equation of motion should pick up a sizable friction term from thermal interactions \cite{Moss:1985wn}. The interaction results in radiation production, which prevents the universe from supercooling. The significance of the extra dissipative term was independently realised in \cite{Yokoyama:1987an} and later in \cite{Berera:1995ie}, where the inflaton dynamics were studied and the scenario was named {\it warm inflation}.  

Over the past several years, warm inflation has been shown to be a valid model of the early universe. The background dynamics of the dissipative inflaton has been modeled \citep{Berera:1996nv, Berera:1997fm} and the non-equilibrium thermodynamical problem has been studied extensively \citep{Berera:1998gx,Berera:1998px,Yokoyama:1998ju,Berera:1999ws,Moss:2000jj,Berera:2001gs,Lawrie:2002wm}. 
The characteristic friction term has been calculated for supersymmetric models with a two stage decay process \cite{Berera:2002sp,Moss:2006gt}. Cosmological implications of dissipative terms in hybrid inflationary models have been extensively studied~\cite{BasteroGil:2004tg,BasteroGil:2006vr}.  Since the characteristic friction terms are not limited to finite temperature effects, the more general name {\it dissipative inflation} shall be adopted here.  

A study of the cosmological perturbations produced by dissipative inflation has identified interesting characteristics in the cosmic microwave background (CMB) power spectra from warm inflation \cite{Hall:2003zp}; a small level of running of the scalar spectral index is predicted, while temperature dependent coupling between the field perturbations can lead to oscillatory spectra. Another distinct characteristic of these models is the predicted level of non-Gaussianity, which is high~\cite{Gupta:2002kn,Gupta:2005nh,Moss:2007cv,Moss:2007qd}. 

A naive best-fit parameter search has been previously completed using the first year WMAP CMB data for one model \cite{Hall:2004ab}, demonstrating the existence of parameters which map this model of warm inflation to cosmological data. This paper aims to extend the previous work, by (a) studying several models of dissipative inflation, (b) using robust Markov Chain Monte Carlo (MCMC) techniques to constrain the model parameters in a subset of models, and (c) using two datasets which significantly improve upon the previous analysis, namely the WMAP three-year data \cite{Speetal06} and the SDSS Luminous Red Galaxy (LRG) power spectrum data \cite{Tegetal06}.

This paper is organized as follows:  Section \ref{secDissDyn} introduces the dynamics of dissipative inflation, both for the background and perturbations. Section \ref{secFlucts} defines the amplitudes for both quantum and thermal fluctuations.  Analytical approximations for the numerical models considered are derived in Section \ref{sec_analytic}, split into three main regimes; weak ($r\ll1$, $T<H$), thermally weak ($r\ll1$, $T>H$) and strong dissipation ($r\gg1$, $T>H$).
The implications of analytic limits are discussed in Section \ref{sec_implications} and the numerical simulation and cosmological constraints are detailed in Section \ref{sec_numeric}.  The results are finally given in Section \ref{sec_results}.

\section{Dissipative Dynamics}
\label{secDissDyn}
\begin{figure}[t!] 
\begin{center}
\scalebox{0.5}{\includegraphics{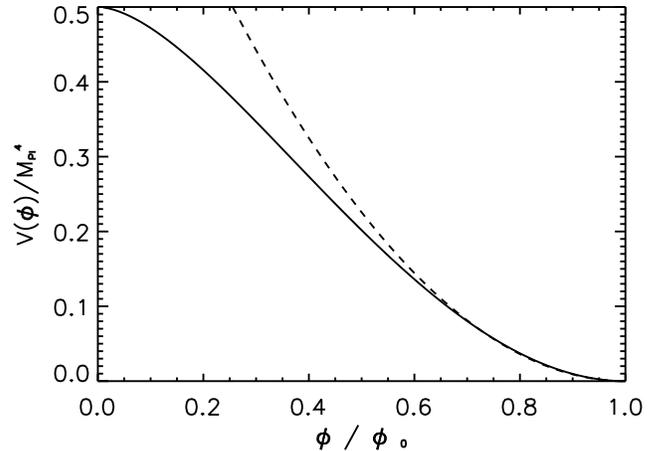}} 
\caption{\label{potential}The full SUSY breaking potential (solid line) used for numerical simulation and the $m^2\phi^2$ potential (dotted line) used in approximation.  Here $m=0.95\mu=0.95\mpl$.}
\end{center}
\end{figure}

The dynamics of dissipative inflation are described in \cite{Hall:2004ab}, where a full discussion of the background and perturbation solutions can be found.  In an expanding, homogeneous universe, the dissipative inflaton equation of motion is given by
\begin{equation}
\ddot\phi+(3H+\Gamma)\dot\phi+V_{,\phi}=0,
\label{wip}
\end{equation}
where
$V(\phi,T)$ is the thermodynamic potential and $\Gamma(\phi,T)$ is the damping term due  to interactions between the inflaton $\phi$ and surrounding fields. Both the thermal correction to the inflaton potential and the damping force are directly due to the effect of the interactions. Issues regarding thermal corrections lead to some debate as to the viability of warm inflation \cite{yokoyama99}.  The corrections to the effective potential were calculated for one explicit model and were shown to be negligible \cite{Hall:2004zr}.  Such corrections shall be ignored for the purpose of this study, in which we will consider a SUSY breaking potential of the form
\begin{equation}
V(\phi)=\frac12\mu^2\left[ \phi^2\log\left({\phi^2\over\phi_0^2}\right)
+\phi_0^2-\phi^2\right].
\label{fullV}
\end{equation}
The form of this potential is shown in Figure (\ref{potential}).

Accounting for the radiation density, $\rho_r$, for a homogeneous scalar field, the total density and pressure, $\rho$ and $p$, can be written
\begin{eqnarray*}
\rho&=& {\textstyle \frac{1}{2}} \dot \phi^2 + V(\phi) + \rho_r \\
p &=& {\textstyle \frac{1}{2}} \dot \phi^2  - V(\phi) + {\textstyle \frac{1}{3}}\rho_r
\end{eqnarray*}
and the zero curvature Friedmann equation becomes
\begin{equation}
3H^2=\frac{1}{\mpl^2} \left[ \frac{1}{2}{\dot\phi}^2+V(\phi) + \rho_r\right],
\label{origfried}
\end{equation}
where $\mpl$ is the reduced Planck mass, $2.436\times 10^{18}GeV$.
Due to the damping term, the continuity equation acquires a source term, which indicates radiation production:
\begin{equation}
\dot \rho_r + 4 H \rho_r = \Gamma {\dot\phi}^2.
\label{origcont}
\end{equation}
The temperature of the heat bath can be calculated using $\rho_r = \pi^2 g_{*} T^4/30$, where $g_*(T)$ is the effective particle number.
The ratio of radiation production to expansion rate, 
\begin{eqnarray}
r=\Gamma / 3H
\end{eqnarray}
is a useful parameter to distinguish between regimes of weak or strong dissipation.

It is usual to assume a slowly-rolling inflaton, so that the second derivatives in the equations of motion may be ignored.  In this limit, {\it slow-roll} is determined by a set of parameters:
\begin{eqnarray}
\epsilon&={\mpl^{2}\over 2}\left({V_{,\phi}\over V}\right)^2,\quad
\eta&={\mpl^{2}}\left({V_{,\phi\phi}\over V}\right),\nonumber \\
\beta&={\mpl^{2}}\left({\Gamma_{,\phi}V_{,\phi}\over \Gamma V}\right),\quad
\delta&={TV_{,\phi T}\over V_{,\phi}}.
\label{slowrp}
\end{eqnarray}
The first three are required to be less than $1+r$.

In the zero-shear, longitudinal gauge,
\begin{eqnarray*}
ds^2=-(1-2\varphi)dt^2+a^2(1+2\varphi)\delta_{ij}dx^idx^j,
\end{eqnarray*}
the perturbed Einstein equations lead to
\begin{eqnarray}
3H\dot\varphi+\left(3H^2+k^2a^{-2}\right)\varphi&=&\frac{\delta\rho}{2\mpl^2}, \nonumber \\
-\ddot\varphi-4H\dot\varphi-\left(2\dot H+3H^2\right)\varphi&=&\frac{\delta p}{2\mpl^2},
\label{pert_metric}
\end{eqnarray}
with
\begin{eqnarray*}
\delta\rho&=&\dot\phi\,\delta\dot\phi+V_{,\phi}\delta\phi
+\dot\phi^2\varphi+ \delta \rho_r, \\
\delta p&=&\dot\phi\,\delta\dot\phi-V_{,\phi}\delta\phi
+\dot\phi^2\varphi+ \frac{1}{3}\delta \rho_r.
\end{eqnarray*}
In addition, the perturbed equation of motion becomes
\begin{eqnarray}
\begin{split}
\delta\ddot\phi+(3H+\Gamma)\delta\dot\phi+\dot\phi(\delta\Gamma) &
+k^2a^{-2}\delta\phi+\delta V_{,\phi}+
4\dot\phi\dot\varphi\\
&-\Gamma\dot\phi\varphi-2V_{,\phi}\varphi=0.
\end{split}
\label{pert_phi}
\end{eqnarray}
Note that when $\Gamma\equiv\Gamma(\phi,T)$, the $\dot\phi(\delta\Gamma)$ term leads to a coupling of the perturbed radiation and inflaton fields.
This non-trivial coupling is important in determining the shape of the power spectrum of perturbations.
Models with inverse-temperature dependent friction terms have been shown to produce oscillations in the power spectrum, with diminishing amplitude for increasing wavenumber\cite{Hall:2003zp}.  
While these models are phenomenologically interesting, we note that, in these models, each oscillation contains a minimum at ${\cal P}_k=0$, which result in zeroes in the power spectrum.  
Since these are not seen in the acoustic spectra, we assume they cannot reproduce observations. 
Further work is required to reduce the amplitude of such oscillations in these models in order to convincingly match to data. 
In addition, friction terms with positive temperature dependence may not lead to oscillation.
These models are beyond the scope of this paper, but is the topic of on-going research.  In this work, we will restrict the form of the friction term to $\Gamma=\Gamma(\phi)$.

One can consider the gauge-invariant variable \citep{Lukash:1980iv,lukash80-1,Bardeen:1980kt}
\begin{eqnarray}
{\cal R}=\varphi-k^{-1}aHv\label{defphi},
\end{eqnarray}
where $v$ is the velocity perturbation.
In the longitudinal (zero-shear) gauge, this variable is equivalent to the {\em curvature perturbation} in the comoving gauge \citep{Lyth:1984gv}.  
Well outside the horizon, $k\ll aH$,
\begin{eqnarray}
{\cal R}=\varphi+\frac13\frac{\delta\rho}{\rho+p}\approx -\frac{H\delta\phi}{\dot\phi}
\end{eqnarray}
Typically, $\cal R$ quickly freezes outside of the horizon and the final amplitude can be approximated using horizon-crossing field values.
The power spectrum for the curvature perturbation is defined by
\begin{eqnarray}
{\cal P_R}(k)(2\pi k^{-1})^3\delta^{(3)}(k-k')&=& \langle R_k R_{-k'}\rangle \\
&\approx& \left[ \frac{H^2}{\dot\phi^2}\langle \delta\phi_k \delta\phi_{-k'}\rangle \right]_*
\label{P_R}
\end{eqnarray}
where the final equation is calculated at horizon crossing.

\section{Fluctuation Amplitudes}
\label{secFlucts}
In supercooled inflation, fluctuations of the inflaton field are seeded by vacuum fluctuations inside the horizon.  The amplitude is well known \cite{Vilenkin:1982wt,Linde:1982uu,Starobinsky:1982ee} and given by
\begin{eqnarray}
\delta\phi(k)=\frac{H}{2\pi}.
\label{quantum}
\end{eqnarray} 
In dissipative inflation when $T>H$, due to the thermal bath, thermal fluctuations in the fields are expected to dominate over their quantum counterparts.  
For the strong dissipative regime, $r\gg 1$, the amplitudes are calculated in \cite{Berera:1999ws,Hall:2003zp}.  In the next section, however, it will be important to consider cases where $r\approx 1$.
The fluctuation amplitudes (with $\Gamma$ consistently replaced by $3H(1+r)$) are given by
\begin{eqnarray}
\delta\phi(k)&\sim&\left({\frac{3\pi}{4}}\right)^{1/4}\left(HT\right)^{1/2}(1+r)^{1/4} \label{dphi}, \\
\delta\dot\phi(k)&\sim& -\left({k \over a}\right)^2 \frac{1}{3H(1+r)} \delta\phi(k). \label{dphidot} 
\end{eqnarray}
As in \cite{Hall:2003zp}, the sign of $\delta\dot\phi$ in Eqn. (\ref{dphidot}) has been chosen for consistency with cross-correlations. Notice that for the strong regime, $r\gg 1$, these results return the values in \cite{Hall:2003zp}.  
In contrast with supercooled inflation, thermal fluctuations do not freeze at Hubble crossing, but at a time given by $ka^{-1}=(\Gamma H)^{1/2}$ called {\it thermal freezeout}~\cite{Berera:1999ws}.  When $r>1$, the freezeout time always precedes the Hubble crossing
time, at which $ka^{-1}=H$. We note that freezeout has been calculated more recently with an additional factor of $\sqrt{3/2}$~\cite{Moss:2007cv}, but for the numerics we have not included this pre-factor and it does not affect our conclusions.

It should be noted that, even for $r<1$, thermal fluctuations may dominate over quantum ones, if $T>H$. 

The fluctuations of the radiation field remains as in \cite{Hall:2003zp}:
\begin{eqnarray}
\delta\rho_r(k)&=&\left({2\pi^2\over
15}\right)^{1/2}\left(k\over a\right)^{3/2}g_*^{1/2}T^{5/2}.\label{drho}
\end{eqnarray} 

Using the above terms for $\delta\phi$ in Eqn.~(\ref{P_R}), we may calculate the spectral index, $n_s-1=d{\cal P_R}/d\ln{k}$ and running of this index, $dn_s/d\ln{k}$~\cite{Liddle:1992wi,Hall:2003zp}, when $\Gamma=\Gamma(\phi)$.
We require two second-order slow-roll parameters:
\begin{equation}
\zeta^2={ \mpl^{4}}\left({V_{,\phi}V_{,\phi\phi\phi}\over
V^2}\right),\quad
\gamma={\mpl^{2}}\left({\Gamma_{,\phi\phi}\over \Gamma}\right).
\label{soslowrp}
\end{equation}
For quantum fluctuations, using Eqn.~(\ref{quantum}), corrections due to the additional friction are found ($r<1$):
\begin{eqnarray}
n_s -1 &=& -6\epsilon_*+2\eta_*+r_*\left(8\epsilon_*-2\eta_*-2\beta_*\right) \\
\frac{dn_s}{ d\ln{k}} &=& 16\epsilon_*\eta_*-24\epsilon_*^2-2\zeta_*^2 
+r_*\left(64\epsilon^2_*+4\gamma_*\epsilon_* \right. \nonumber \\
&& \qquad \left. -14\epsilon_*\beta_*-38\eta_*\epsilon_* +4\beta_*\eta_*+4\zeta_*^2\right) \nonumber
\end{eqnarray}
For thermal fluctuations when $r<1$, using Eqn.~(\ref{dphi}):
\begin{eqnarray}
n_s -1 &=& 
-\frac{15}{4}\epsilon_*+\frac{3}{2}\eta_*-\frac{3}{4}\beta_* \\
&& + r_*\left(\frac{21}{4}\epsilon_*-\frac{3}{4}\beta_*-\frac{3}{2}\eta_*\right)\nonumber \\
\frac{dn_s}{ d\ln{k}} &=& -15\epsilon_*^2+\frac{21}{2}\epsilon_*\eta_*-\frac32\epsilon_*\beta_*+\frac32\epsilon_*\gamma_*+\frac{3}{4}\beta_*\eta_* \nonumber \\
&& -\frac{3}{4}\beta_*^2-\frac32\zeta_*^2 +r_*\left(\frac{165}{4}\epsilon_*^2+6\epsilon_*\beta_* \right. \nonumber \\
&& \left. -\frac{51}{2}\epsilon_*\eta_*+\frac32\beta_*\eta_*+\frac34\beta_*^2+3\zeta_*^2\right) \nonumber
\end{eqnarray}
When $r>1$, thermal fluctuations lead to:
\begin{eqnarray}
n_s &=& {1\over r_* }\left(-{9\over 4}\epsilon_*+{3\over 2}\eta_*-
{9\over 4}\beta_*\right) \\
\frac{dn_s}{ d\ln{k}} &=& {1\over r_*^2
}\left(-\frac92\beta_*^2-\frac{27}4\epsilon_*^2-\frac92\epsilon_*\beta_*
 +\frac{15}4\eta_*\beta_* \right. \nonumber \\
&& \qquad \qquad \qquad\left. +6\epsilon_*\eta_*-\frac32\zeta_*^2+\frac92\gamma\epsilon_*
\right). \nonumber
\end{eqnarray} 
Due to the constancy of the curvature perturbation outside the horizon, the spectral index may be calculated at horizon crossing or freezeout, as indicated by an asterisk.

The spectrum of gravitational waves is given by~\cite{Liddle:1992wi}
\begin{eqnarray}
{\cal P}_g=\frac{2V}{3\pi^2\mpl^4}
\label{gravwaves}
\end{eqnarray}
If fluctuations in the radiation field exist, then isocurvature fluctuations should be expected.  Their spectrum is given by~\cite{Taylor:2000ze}:
\begin{eqnarray}
{\cal P}_{\rm iso}=\frac{\left(\Gamma H\right)^{1/2}T}{400\pi^3\mpl^2}.
\label{iso}
\end{eqnarray}
Note that the difference in prefactors with respect to~\cite{Taylor:2000ze} arises from the definition of the scalar perturbation amplitude.
The scalar-to-tensor ratio and ratio of isocurvature perturbations can be defined by
\begin{eqnarray}
R_g=\frac{{\cal P}_g}{{\cal P_R}}, \qquad R_{\rm iso}=\frac{{\cal P}_{\rm iso}}{{\cal P_R}}.
\end{eqnarray}

\section{Analytical Approximation}
\label{sec_analytic}
In the next section, we will numerically evolve the full equations of motion, solving for both the background and perturbations.  The full system is complicated and the results are non-intuitive.
It is therefore useful to find an approximate analytic solution, which should provide a good guide for the expected parameter ranges. 
The full potential is the SUSY breaking one given in Eqn. (\ref{fullV}).  Near the minimum, it is possible to consider a shifted $m^2\phi^2$ potential:
\begin{eqnarray}  
V(\phi) = m^2\left(\phi_0-\phi\right)^2,
\end{eqnarray}  
which approximates closely to the full potential close to the minimum (see Fig \ref{potential}).  A good fit occurs when $m\approx0.95\mu$.
In addition, we shall restrict ourselves to the simple case in which
\begin{eqnarray}
\Gamma(\phi)=\Gamma_0 \left(\frac{\phi}{\phi_0}\right)^n.
\end{eqnarray}

The slow-roll regime can be assumed, for which the equations of motion simplify:
\begin{eqnarray}
\dot\phi=-\frac{V^{\prime}}{3H+\Gamma}, \quad \quad \rho_r = \frac{\Gamma\dot\phi^2}{4H},
\end{eqnarray}
and the slow-roll parameters are given by
\begin{eqnarray}
\epsilon=\eta=\frac{2\mpl^2}{\left(\phi_0-\phi\right)^2}, \quad \quad \beta=\frac{\mpl^2}{\phi\left(\phi_0-\phi\right)},  \quad \quad \delta=0.
\end{eqnarray}

For this potential, the index is given by:
\begin{eqnarray}
1-n_s \approx
\left\{ \begin{array}{ll}
\epsilon_*(4 -6r) & r<1 ~(\mathrm{quantum}) \nonumber \\
& \nonumber \\
\epsilon_*\left(\frac{9}{4}-\frac{15}{4}r_*\right)& r<1 ~(\mathrm{thermal}) \\
& \nonumber \\
\frac{3}{4}\frac{\epsilon_*}{r_*}& r> 1 ~(\mathrm{thermal})
\end{array} \right. 
\label{indices}
\end{eqnarray}
where we have assumed that $\beta$ is negligible compared to $\epsilon$ and $\eta$ (which is really only true as $\phi\rightarrow\phi_0$).  Notice that for all three cases, a {\it red} index is expected.
Due to the dominance of $\epsilon$ (and $\eta$), we expect the running to be $O((1-n_s)^2)$, hence of the order of $10^{-3}$ or less.

\subsection{Background Values}
\label{sec_background}
In the weakly dissipative limit, $1-n_s\approx4\epsilon_*$ leads to
\begin{eqnarray}
\frac{\phi_*}{\phi_0} = 1 - \sqrt{\frac{8}{1-n_s}}\frac{\mpl}{\phi_0}.
\label{weakphistar}
\end{eqnarray}
Conversely, in the strong limit,
\begin{eqnarray}
\phi_*^n(\phi_*-\phi_0) = \frac{6\sqrt{3}\mpl\phi_0^n m}{(1-n_s)\Gamma_0}=C.
\label{strongC}
\end{eqnarray}
For large $\Gamma_0$, $C$ is small and, for all $n$, it is clear that either 
\begin{eqnarray}
\phi_*\approx0,\qquad \phi_*\approx\phi_0.
\label{strongphistar}
\end{eqnarray}
We are interested in the case in which inflation occurs very close to $\phi_0$ and hence we take this limit.
Note that, in this regime, the friction is very large and hence a large number of $e$-foldings may be obtained even when the inflaton is close to its minimum.

At the end of inflation, $\epsilon\approx(1+r)$.  In fact, for strong dissipation, $\epsilon\approx2r$ \cite{Taylor:2000jw}. Hence, in these two limits,
\begin{eqnarray}
\phi_f|_{r<1} &=& \phi_0 - \sqrt{2}\mpl , \\
(\phi_0-\phi_f)\phi_f^n |_{r>1} &=& \frac{\sqrt{3}m\mpl\phi_0^n}{\Gamma_0}.
\label{phif}
\end{eqnarray}
This agrees with the previous results, Eqn. (\ref{weakphistar}) and (\ref{strongphistar}), where $\phi_f\approx\phi_0$ in both limits.

\begin{figure*}[t!]
\begin{center}
\scalebox{0.5}{\includegraphics{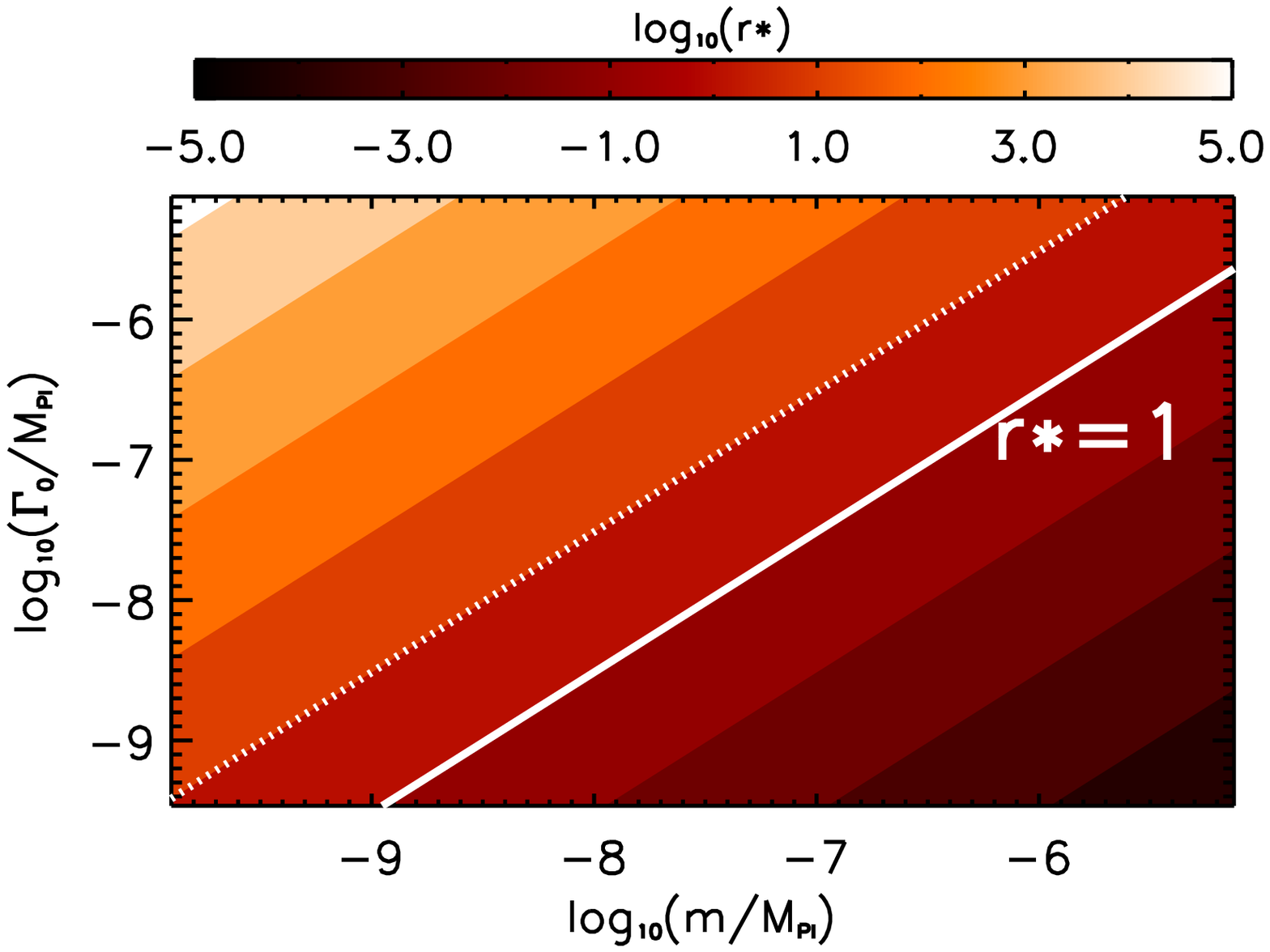}}
\scalebox{0.5}{\includegraphics{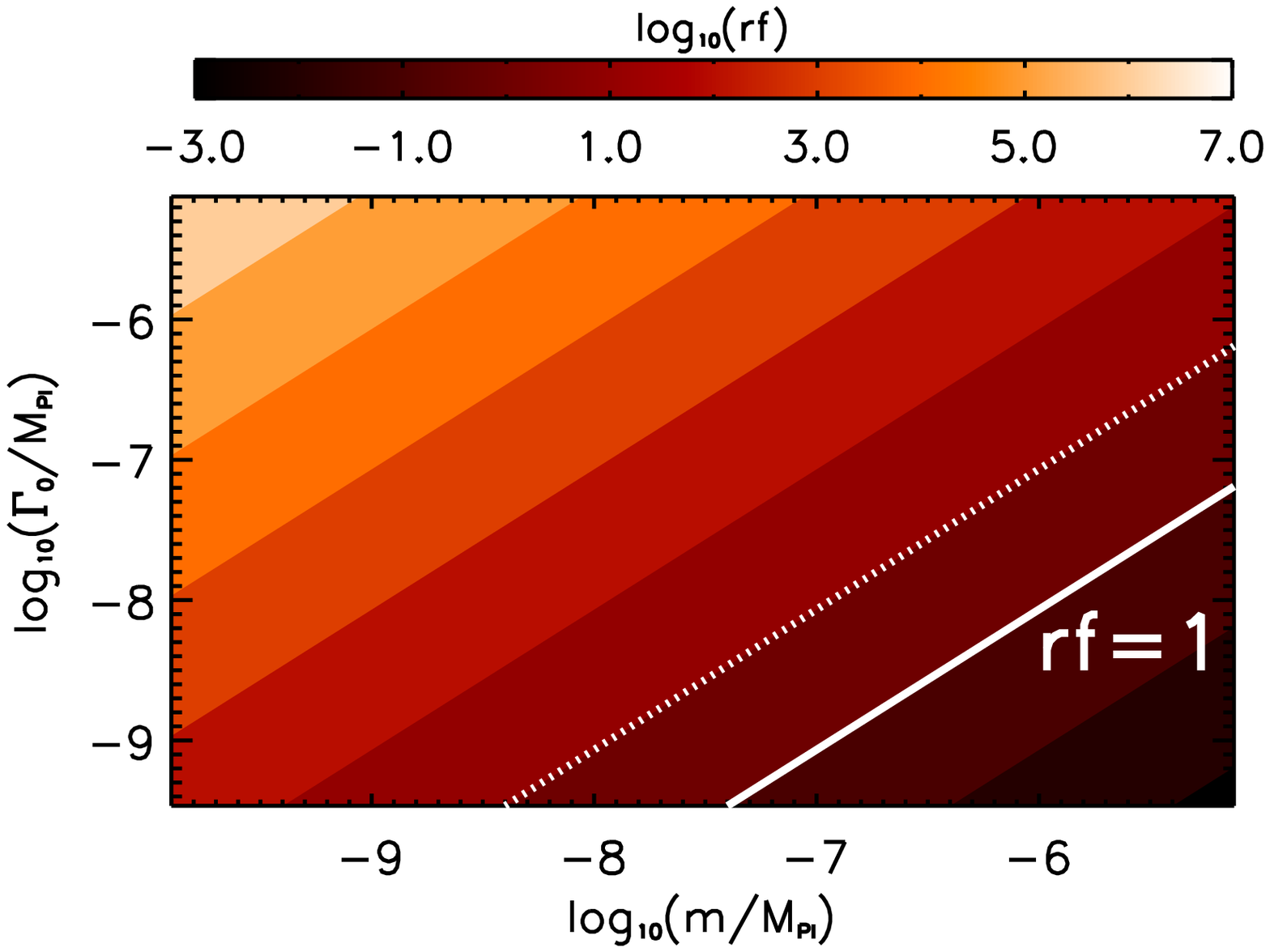}}
\caption{\label{Fig2}The ratio of the friction terms at horizon crossing ($r_*$, LHS plot) and end of inflation ($r_f$, RHS plot), which determines weak ($r<1$) and strong ($r>1$) regimes. A value of $\phi_*=0.8\phi_{0}$ is assumed in Eqn. (\ref{rstar}) as detailed in the text. A value of $n=1$ has been assumed and $\phi_0=\mpl$ (solid line).  For reference, the line $r_*=1$ when $\phi_0=10\mpl$ has also been plotted (dotted line). }
\end{center}
\end{figure*}

It is convenient to plot the dependence of $r$ on $\Gamma_0$ and $m$, in order to determine the weak and strong regimes.  
For this potential,
\begin{eqnarray}
r = \frac{\Gamma}{3H} = \frac{\Gamma_0\mpl}{\sqrt{3}~m} \frac{\phi^n}{\phi_0^n\left(\phi_0-\phi\right)}.
\label{rstar}
\end{eqnarray}
This ratio at horizon-crossing and at the end of inflation, $r_*$ and $r_f$, is dependent on $\phi_{*}$ or $\phi_{f}$ respectively, which in turn depend on the parameters $\Gamma_0$, $m$, and $\phi_0$, as in Eqns (\ref{strongC}-\ref{phif}).  
From Eqns (\ref{weakphistar}) and (\ref{strongphistar}), we may take an average value of $\phi_*\approx 0.8\phi_{0}$.  For $\phi_f$ we solve Eqn. (\ref{phif}).  
Results for $r_*$ and $r_f$ for relevant ranges of $\Gamma_0$ and $m$ are plotted in Figure (\ref{Fig2}) for $n=1$.

\begin{figure}[t!]
\begin{center}
\scalebox{0.5}{\includegraphics{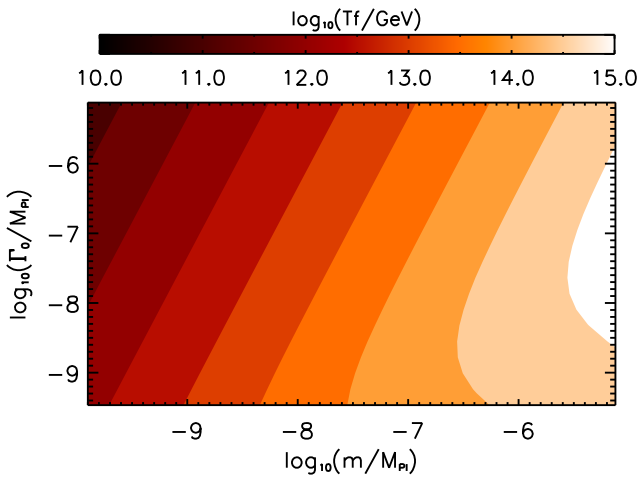}}
\caption{\label{FigTf}The final reheat temperature, $T_f$, over the full parameter range considered. A value of $n=1$ has been assumed and $\phi_0=\mpl$.}
\end{center}
\end{figure}

The radiation density can be calculated using the slow-roll approximation~\cite{Taylor:2000jw},
\begin{eqnarray}
\rho_r\approx\frac{3}{4}r\dot\phi^2\approx\frac{r~V\epsilon}{2(1+r)^2}=\frac{r~m^2\mpl^2}{(1+r)^2},
\label{rhoeqn}
\end{eqnarray}
from which the temperature can be approximated.  Substituting $r_f$ into this equation leads us to the final reheat temperature, plotted in Figure (\ref{FigTf}) for $n=1$.  
When $n>1$, the final values $r_f$ and $T_f$ do not differ much from the $n=1$ case, since $\phi_f\approx\phi_0$.  

From Eqn.~(\ref{rhoeqn}), it is also possible to approximate the ratio $T/H$:
\begin{eqnarray}
\frac{T}{H}=\frac{\sqrt{3}}{\left(2\alpha\right)^{1/4}}\frac{r^{1/4}\epsilon^{1/4}}{(1+ r)^{1/2}}
\sqrt{\left(\frac{\mpl}{\phi_0}\right)\left(\frac{\mpl}{m}\right)\left(1-\frac{\phi}{\phi_0}\right)^{-1}}
\end{eqnarray}
Note that when $r\ll1$, $T/H \propto r^{1/4}$ and the ratio is suppressed by $r$ (although increases with $r$). When $r\gg1$, $T/H \propto r^{-1/4}$, but since $\epsilon\approx(1-n_s)r$, the dependence on $r$ cancels.

Finally it is possible to estimate the number of $e$-foldings during inflation:
\begin{eqnarray}
N = \int_{\phi_*}^{\phi_f} \frac{H}{\dot\phi}d\phi 
\approx-\frac{1}{\mpl^2} \int_{\phi_*}^{\phi_0} \frac{V(1+r)}{V^{\prime}}d\phi,
\end{eqnarray}
resulting in 
\begin{eqnarray}
N\approx
\left\{ \begin{array}{ll}
\frac{1}{4\mpl^2}
\left(\phi_0-\phi_*\right)^2 & r\ll 1 \\ 
& \\ 
\frac{\Gamma_0}{4\sqrt{3}m\mpl\phi_0^n(n+1)}
\left(\phi_0^{n+1}-\phi_*^{n+1}\right) & r\gg 1
\end{array} \right.
\label{efoldings}
\end{eqnarray}

In all the cases we shall now consider, the scalar-to-tensor ratio is set by the scale of the potential:
\begin{eqnarray}
R_g\approx\frac{2m^2\phi_0^2}{3\pi^2 {\cal P_R} \mpl^4}.
\end{eqnarray}
A very conservative upper limit is found when $m/\mpl\sim 10^{-5}$ and $\phi_0/\mpl\sim 10$ (values considered later in the paper), such that 
\begin{eqnarray}
R_g<0.2. \nonumber
\end{eqnarray}

\subsection{Weak Dissipation Amplitudes}
\label{SecWeak}
\begin{figure*}[t!]
\begin{center} 
\scalebox{0.5}{\includegraphics{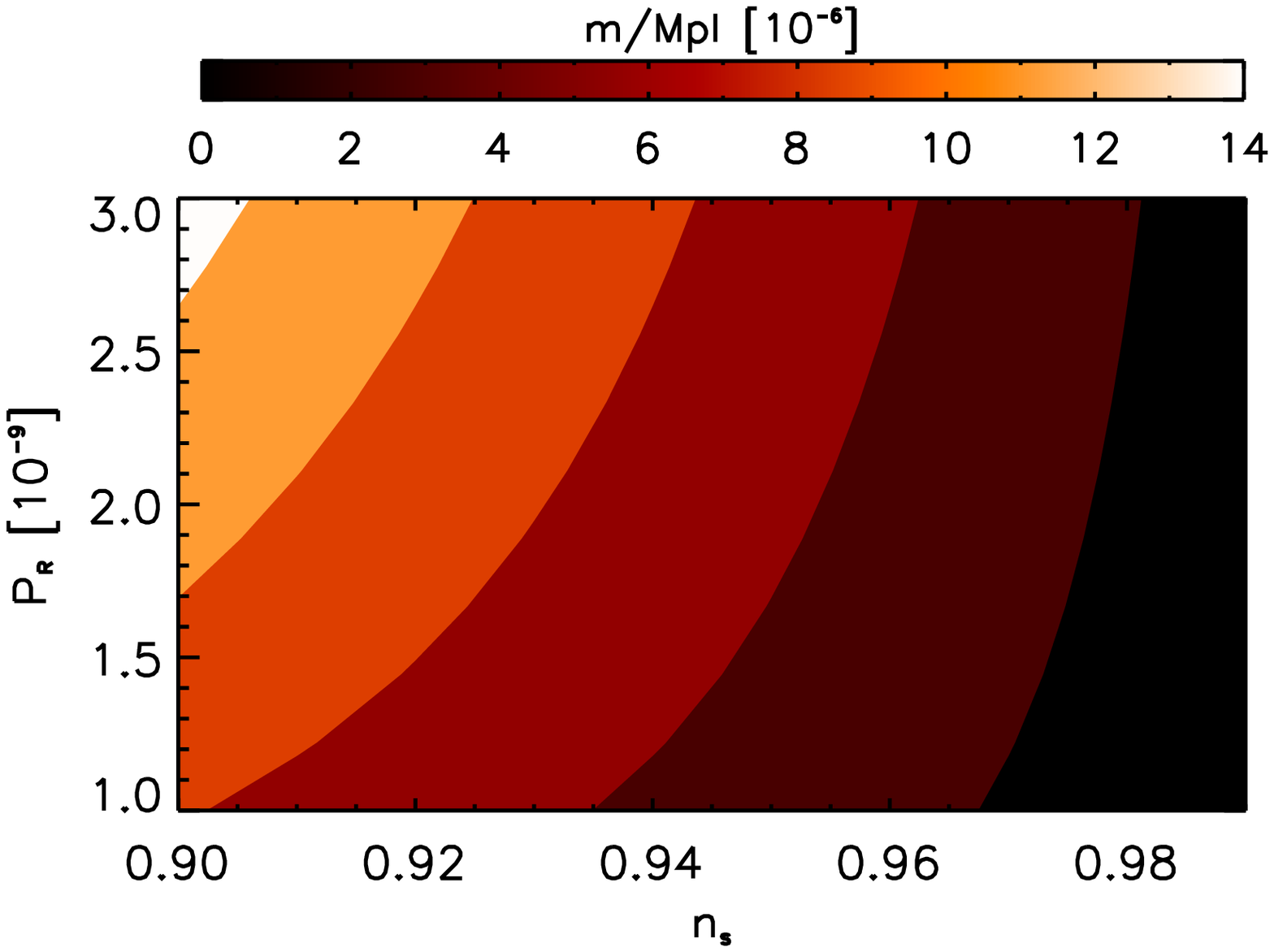}} 
\scalebox{0.5}{\includegraphics{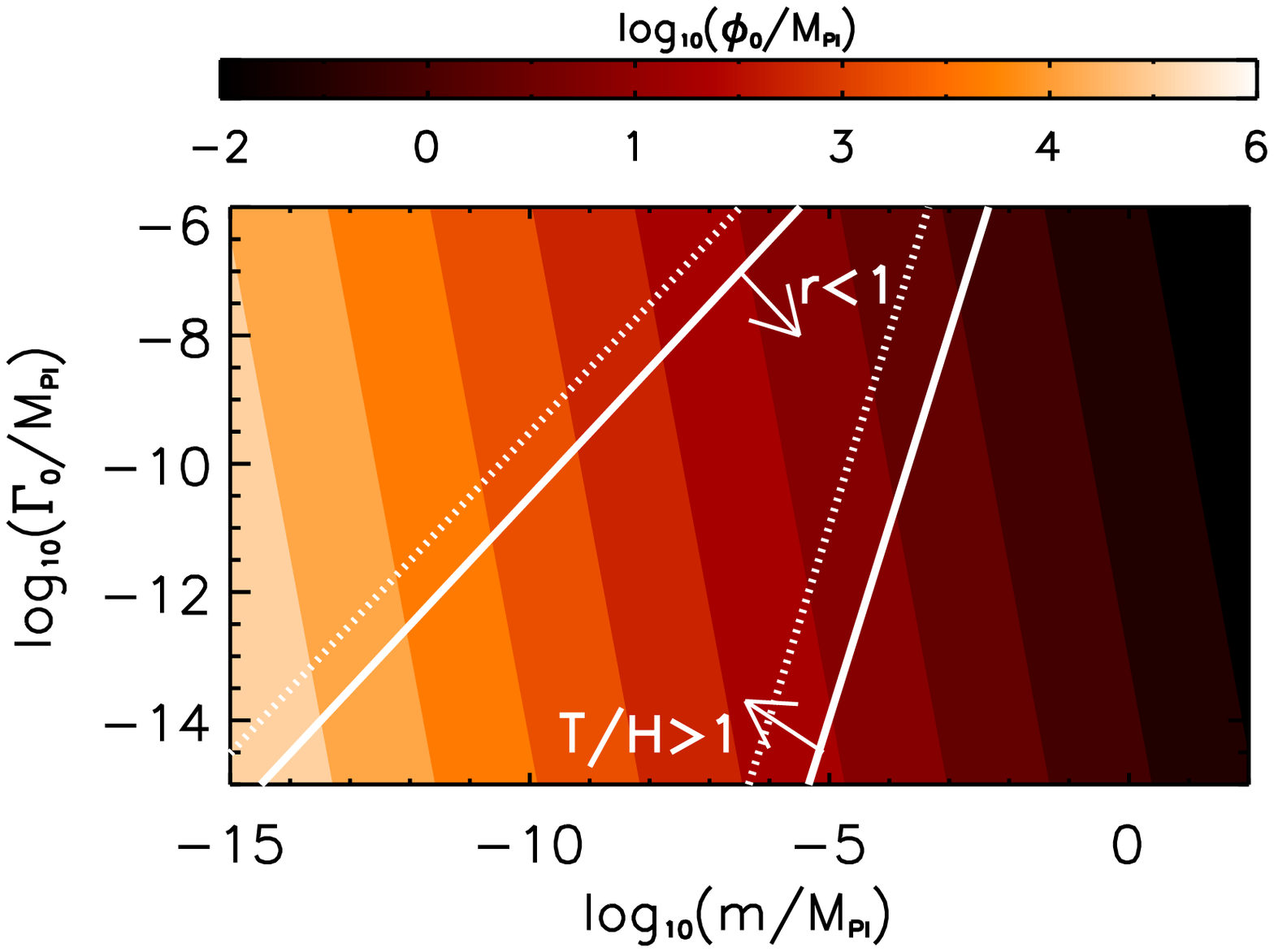}}
\caption{\label{weak_constraint}Expected parameter values in the weak regime ($r\ll 1$) when $n=1$.  When $T<H$ (LHS plot), the scale of the potential, $m$, is constant, given in terms of the amplitude $\cal P_R$, and index $n_s$ as in Eqn.~(\ref{analyticm}).  When $T>H$ (RHS plot), the three parameters are related by Eqn.~(\ref{thermalweak}) and are plotted for typical WMAP amplitudes.  In addition, the constraints for $r<1$ and $T>H$ are also shown for $\phi_0=\mpl$ (solid) and $\phi_0=10\mpl$ (dotted).  Note that when $T>H$, a further constraint comes from $N>60$ (or $\phi_0\gtrsim\mpl$).}
\end{center}
\end{figure*}

In the weak dissipative limit, the quantum fluctuations dominate and the amplitude of fluctuations is given by Eqn. (\ref{quantum}).  In this case, the amplitude of perturbations at horizon crossing is
\begin{eqnarray}
{\cal P_R} &=& \frac{H^4}{4\pi^2 \dot\phi^2} \\
           & =& \frac{1}{24\pi^2}\frac{V}{\mpl^4}\frac{1}{\epsilon_*}\\
           &=& \frac{1}{12\pi^2}\frac{m^2}{\mpl^2}\frac{1}{\epsilon_*^2}.
\end{eqnarray}
Substituting $\epsilon_*\approx(1-n_s)(1+\frac32r)/4$, we find:
\begin{eqnarray}
\frac{m}{\mpl}&=&\left(\frac{3{\cal P_R}}{4}\right)^{1/2}\pi(1-n_s)(1+\frac32r) \nonumber \\ 
                &=&2.72~{\cal P_R}^{1/2}(1-n_s)(1+\frac32r).
\label{analyticm}
\end{eqnarray}
Note that the result is independent of $n$, since the amplitude is independent of $\Gamma$ (if we assume $r$ is negligible).
The range of values allowed given ${\cal P_R}$ and $n_s$ is shown in Figure (\ref{weak_constraint}).  For typical WMAP values, $m\approx10^{-6}\mpl$.  
We note that the scalar-to-tensor ratio, $R_g$, is calculated using Eqn.~(\ref{gravwaves}) and is given by the standard supercooled limit of $R_g=16\epsilon$.
The ratio of isocurvature perturbations is given by
\begin{eqnarray}
R_{\rm iso} &=& \left(\frac{9}{2\alpha}\right)^{1/4}\frac{r^{3/4}\epsilon^{5/4}}{50\pi}\frac{\mpl}{V^{1/4}} \nonumber \\
&\approx& 6 \times 10^{-3}\left(\frac{\Gamma}{\mpl}\right)^{3/4}\left(\frac{\mpl}{m}\right)^{5/4}\left(\frac{\mpl}{\phi_0}\right)^{15/4} \nonumber
\end{eqnarray}

\subsection{Thermally Weak Amplitudes}
In this regime, the friction term acts only weakly $r<1$, but the thermal amplitude dominates over its quantum counterpart, such that the amplitude is given by:  
\begin{eqnarray}
{\cal P_R} &=& \left(\frac{3\pi}{4}\right)^{1/2}\frac{H^{3}T}{\dot\phi^2} \nonumber \\
        &=& \frac{\pi^{1/2}}{2^{9/2}3^{1/8}\alpha^{1/4}}\frac{V^{5/8}\Gamma^{1/4}}{\epsilon_{*}^{3/4}\mpl^{11/8}} \nonumber \\
        &\approx& 0.0285~\frac{m^{5/4}\phi_0^{11/4}\left(1-\frac{\phi_{*}}{\phi_{0}}\right)^{11/4}\Gamma^{1/4}}
{\mpl^{17/4}}.
\label{thermalweak}
\end{eqnarray}
For $r<1$, from Eqn.~(\ref{rstar}) we require:
\begin{eqnarray}
\frac{1}{\sqrt{3}}\frac{\Gamma}{\mpl}\frac{\mpl}{m}\frac{\mpl}{\phi_{0}}\left(1-\frac{\phi_{*}}{\phi_{0}}\right)^{-1} <1
\end{eqnarray}
and imposing $T/H>1$ (with $r<1$), we find
\begin{eqnarray}
\left(\frac{\Gamma}{\mpl}\right)\left[\frac{\mpl}{m}\frac{\mpl}{\phi_{0}}\left(1-\frac{\phi_{*}}{\phi_{0}}\right)^{-1}\right]^{3} >9\times 10^4 \left(1-n_s\right).
\end{eqnarray}
These constraints on the parameters $m$ and $\Gamma_0$ are shown in Fig.~\ref{weak_constraint}, when $\phi_0/\mpl=1,10$.  Also superimposed are the required values of $\phi_0$ which satisfy Eqn.~(\ref{thermalweak}) for typical WMAP values.
The ratio of isocurvature perturbations is given in this regime by
\begin{eqnarray}
R_{\rm iso} = \sqrt{\frac{3r}{\pi^7}}\frac{\epsilon}{100}. \nonumber
\end{eqnarray}

\subsection{Strong Dissipation Amplitudes}
\label{SecStrong}
For the perturbations in this regime, the new friction term dominates ($r\gg 1$) and, assuming $T>H$, the amplitude is given by:
\begin{eqnarray}
{\cal P_R} &=& \left(\frac{\pi}{4}\right)^{1/2}\frac{H^{5/2}\Gamma^{1/2}T}{\dot\phi^2} \label{approxGamma} \\
           &=& \left(\frac{\pi}{4}\right)^{1/2}\frac{1}{(4\alpha)^{1/4}}
                 \left(\frac{\Gamma}{\mpl}\right)^{3/2}
                 \left(\frac{r}{6\epsilon_*}\right)^{3/4}\nonumber  \\
&=& \left(\frac{\pi^2}{64\alpha}\right)^{1/4}
                 \left[\frac{\Gamma_0}{\mpl}\left(\frac{\phi_*}{\phi_0}\right)^n\right]^{3/2}
                 \left[\frac{1}{8(1-n_s)}\right]^{3/4} \nonumber,
\end{eqnarray}
where we have used $\rho_r=\pi^2g_*T^4/30=\alpha T^4$ and $(1-n_s)\approx3\epsilon_*/4r_*$.
There is still some field dependence in this equation, due to the form of $\Gamma(\phi)$, but from Eqn. (\ref{strongphistar}) we see that this dependence is very weak and can be ignored.
For this reason, the dependence on the index $n$ is also very weak.
\begin{figure}[t!]
\begin{center}
\scalebox{0.5}{\includegraphics{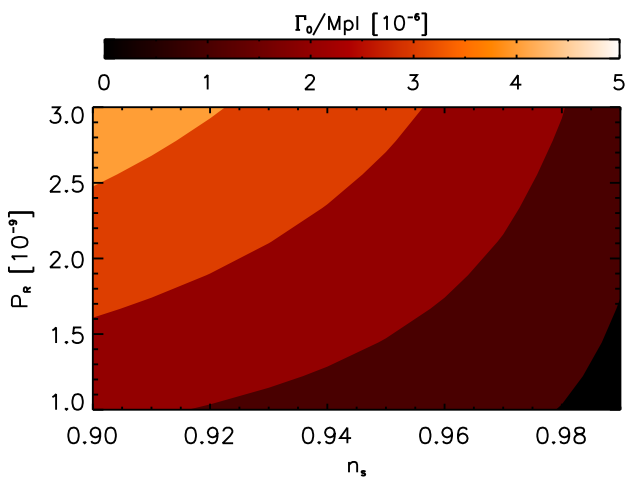}} 
\caption{\label{strong_constraint}Expected parameter values in the strong regime ($r\gg 1$) when $n=1$.  In this strong regime, $\Gamma_0$ is constant and is given in terms of the amplitude $\cal P_R$, and index $n_s$ by Eqn.~(\ref{analyticG}).}
\end{center}
\end{figure}

It is immediately obvious that, in this strong limit, the amplitude in Eqn. (\ref{approxGamma}) is independent of the mass $m$.  In fact,
\begin{eqnarray}
\frac{\Gamma_0}{\mpl}&=& 6.91 {\cal P_R}^{2/3}(1-n_s)^{1/2} \left(\frac{\phi_0}{\phi_*}\right)^n,
\label{analyticG}
\end{eqnarray}
where the last equation has used $g_*=100$ ($\alpha\approx 32.9$).  Figure (\ref{strong_constraint}) shows how $\Gamma_0$ depends on realistic values of $\cal P_R$ and $n_s$ when $n=1$.  For typical WMAP amplitudes, $\Gamma_0\approx10^{-6}\mpl$.
The ratio of isocurvature perturbations in the strong regime is given by~\cite{Taylor:2000ze}:
\begin{eqnarray}
R_{\rm iso} = \frac{1}{100\pi^{7/2}}\frac{\epsilon}{r^2}. \nonumber
\end{eqnarray}

\section{Implications for Particle Physics}
\label{sec_implications}

The analytic results lend some insight into the allowed physical scales of the inflationary potential.
In the weak dissipation regime, as expected for supercooled inflation, the amplitude of the perturbations, as seen by observation, merely sets the scale of the potential for the inflaton field, as seen in Eqn.~(\ref{analyticm}).
Conversely, however, no limits are placed on the scale of the inflaton in the strongly dissipative regime, whilst the amplitude of the friction term is fixed.  

One of the most stringent constraints on these models will be the reheat temperature, which is of order $10^{10}$-$10^{14}$ GeV in these models.  Counter-intuitively, weak dissipation normally leads to a larger reheat temperature.
In order to evade an overproduction of gravitinos, the reheat temperature should be less than $10^{9}$ GeV~\cite{Olive:1989nu,Khlopov:1984pf} (for a review see \cite{Binetruy:2006ad}).  
From Figures (\ref{Fig2}) and (\ref{FigTf}), it is clear that this requires $r\gg10^3$, which in turn requires $\Gamma_0/\mpl\gg10^{-5}$ for weak dissipation or $m/\mpl\ll 10^{-10}$ for strong dissipation.  
Alternatively, one must consider a mechanism to dilute the gravitinos after inflation ends, such as late-time entropy production.

Additionally, due to high levels of predicted non-Gaussianity, extra constraints are placed on the parameters.  From \cite{Moss:2007cv}, the non-linearity parameter, $f_{NL} ^\mathrm{equil} $ (in the equilateral limit) is given by
\begin{eqnarray}
f_{NL} ^\mathrm{equil}\approx-15\ln\left(1+\frac{r}{14}\right)-\frac{5}{2}.
\end{eqnarray}
Present constraints on the non-linearity parameter give $-256 < f_{NL}^\mathrm{equil}<332$ at $95\%$~\cite{Creminelli:2006rz} and result in $r\leq 10^{10}$.  This places a rather loose lower bound on $m$, $\phi_0$ and $\Gamma_0$, as given in Eqn. (\ref{rstar}). 
However, this constraint is expected to improve with data from the Planck satellite; a 1-$\sigma$ limit on non-Gaussianity $\vert f_{NL}^\mathrm{equil}\vert < 66.9$ is forecast for the latter~\cite{Smith:2006ud}.  This corresponds to $r\approx1000$ \cite{Moss:2007qd}.  
If the dissipative regime evades the gravitino problem, the level of non-Gaussianity should be seen by Planck.

\section{Numerical Simulation}
\label{sec_numeric}

Now we will discuss the numerical simulation which forms the basis of our cosmological constraints. It is important to note that we will only use the previously derived analytic results to cross-check the numerical code. No analytic or slow roll approximations are used in the numerical computations of the cosmological observables which are compared with the data.

It is convenient to adopt a general form for the friction term
\begin{eqnarray}
\Gamma(\phi)=\Gamma_0 \left( \frac{\phi}{\phi_0} \right)^n.
\end{eqnarray}
In order to check the above analytical approximations and identify the effect of higher powers of $\phi$, we will consider two numerical cases:
\begin{itemize}
\item {\bf Case I:} $n=1$
\item {\bf Case II:} $n=2$.
\end{itemize}
The case in which $n=0$ (a constant friction term) is unphysical and will be ignored.  

We choose to study both strong and weak dissipative regimes for both cases I and II.
Of course, in the weak regime, there are two separate scenarios: $T>H$ and $T<H$ (``thermally weak'' or ``weak'').
We therefore study a total of six models: weak, thermally weak and strong, for both $n=1,2$.
 
We expect the models to give a flat spectrum of curvature perturbations, with a small, inherent amount of blue to red running \cite{Hall:2003zp,Hall:2004ab}.  Although the running is small, we are interested in how this running leads to a constraint of the parameters.  
The chosen inflaton potential originates from soft supersymmetry breaking and is given by Eqn.~(\ref{fullV}).
Note that slow roll supercooled inflation cannot occur with this potential, since the standard slow-roll conditions are difficult to satisfy.

\subsection{Dissipative Inflation Code}

We consider the evolution of perturbations for both cases I and II in weak, thermally weak and strong dissipative scenarios.
The background dynamics follow Eqns.~(\ref{wip}-\ref{origcont}), while the perturbations are given by Eqns.~(\ref{pert_metric}-\ref{pert_phi}).
The code follows Ref. \cite{Hall:2003zp} with a few modifications, detailed here.  

For $T<H$, the perturbations freeze at the Hubble radius ($k=aH$) with an amplitude given by Eqn.~(\ref{quantum}).  
When $T>H$, thermal fluctuations dominate; freezeout occurs earlier at a time $k=a(\Gamma H)^{1/2}$ and the amplitudes are given by Eqns. (\ref{dphi}-\ref{drho}).
Wavenumbers $k/a$ are normalised by the Hubble length at the end of inflation:
\begin{eqnarray}
\ln\left(\frac{0.002\ \mathrm{Mpc}^{-1}\ c}{a_fH_f}\right) = -\ln\left(g_*^{1/2} T_{14}\right)-53.6
\end{eqnarray}
where $T_{14}\times10^{14}$ GeV is the temperature at the end of inflation.

We separate the weak and strong dissipative scenario from the weakly thermal regime.  
This latter regime leads to a tightly constrained region of allowed parameter space and will be discussed separately in the next section. 

For what follows, we consider the two separate regimes of weak and strong dissipation, $r<1$ and $r>1$, with quantum and thermal fluctuations respectively.
From the analytical constraints in Section~(\ref{sec_analytic}), we may assume numerical priors on the parameters, $\mu$ and $\Gamma_0$:
\begin{displaymath}
r\ll 1
\left\{ \begin{array}{lcl}
-~6.1 &< \log_{10}\left(\mu/\mpl\right) &<  -5.7 \\
& \\
-10.1 &< \log_{10}\left(\Gamma_0/\mpl\right) &< -5.7
\end{array} \right.
\end{displaymath}

\begin{displaymath}
r\gg 1
\left\{ \begin{array}{lcl}
-10.4 &< \log_{10}\left(\mu/\mpl\right) &< -7.1 \\
& \\
-~7.5 &< \log_{10}\left(\Gamma_0/\mpl\right) &<  -5.7
\end{array} \right.
\end{displaymath}

Analytically, there is no constraint on $\phi_0$ from either strong or weak dissipation.  
However, from Section~(\ref{sec_background}), it is clear that $\phi_0$ sets the total number of $e$-foldings. 
For numerical convenience, we impose a prior on $\phi_0$ using the number of $e$-foldings:  
we impose an upper limit of $N<10^5$, which requires $\log_{10}\left(\phi_0/\mpl\right)\lesssim 2$ for both cases I and II (also both strong and weak), which is greatly above normally considered values.

There is a heuristic reason for why $N<10^5$ leads to the same cut-off in both strong and weak cases.  In all the cases considered, the friction term increases monotonically with $\phi$.
Therefore, at the top of the potential (the beginning of inflation), $\Gamma$ is small and we always have a weak dissipative regime.  
From Eqn. (\ref{efoldings}), it may be observed that, at this time, $N\approx (\phi_0/\mpl)^2$.  
Of course, as $r$ increases, the friction term becomes more important earlier and earlier and $N$ will depend on $\Gamma_0$ and $m$ as well as $\phi_0$.

For strong dissipation, imposing a cut-off on $N$ also imposes a prior on $m$ as seen in Eqn. (\ref{efoldings}).  This lower bound will be seen in the results later.

A natural lower prior on $\phi_0$ is imposed due to the requirement that cosmological modes cross the horizon during inflation (i.e. $N\gtrsim 60$). 

As established in Section~\ref{sec_analytic}, the isocurvature and tensor amplitudes for the ranges we consider are highly subdominant to the scalar amplitude and are negligible considering the quality of data we compare to. 
Therefore we do not compute them numerically.

\subsection{Curvature Perturbation across the Horizon}
The curvature perturbation, ${\cal R}$, which defines the amplitude of the power spectrum, was shown to be approximated by $H\delta\phi/\dot\phi^2$.  
Although it is usually assumed that the curvature perturbation does not deviate from its horizon crossing value, there is some time-dependence.  
Due to the additional terms in Eqn.~(\ref{pert_phi}), $\delta\phi$ can oscillate, shifting the final value of the curvature perturbation.
Hence the final value, ${\cal R}_f$ is not always closely approximated by ${\cal R}_*$.

When $r\ll1$, Eqn.~(\ref{pert_phi}) reduces to the supercooled case and the curvature perturbation does not change significantly across the horizon as expected.  
When $r\gg1$, the friction acts to damp the oscillations and again no significant change in ${\cal R}$ is seen.  
However, a slight shift in the above analytical estimates for $m$ and $\Gamma_0$ is expected numerically.

In the thermally weak regime, however, the final curvature perturbation (as calculated numerically) is not well approximated by the freeze-out values.  
In this thermally weak regime, the final amplitude is not given by the expression in Eqn.~(\ref{thermalweak}), although we have numerically checked that the values at freeze-out are given by this expression.  
It is not possible to approximate the time-dependence of $\delta\phi$ and instead we empirically fit the final amplitudes to an assumed form.  Numerically, we predict a form
\begin{eqnarray}
{\cal P_R}\approx {\cal A}\left(\frac{m}{\mpl}\right)^{\cal B}\left(\frac{\phi_0}{\mpl}\right)^{\cal C}\left(\frac{\Gamma_0}{\mpl}\right)^{\cal D}
\label{empirical}
\end{eqnarray}
where the coefficients are given in Table~\ref{tab_thermalweak} for $n=1,2$.
Due to the small numerical range of $\phi_0$ considered numerically, $0<\log_{10}(\phi_0/\mpl)<2$, the exact dependence on $\phi_0$ is difficult to quantify.
However, note that ${\cal P_R}^{2/3}\propto \left(m^2/\Gamma_0\right)$.
Therefore, for a given value of $\phi_0$, the region of allowed parameter space in the $\log(m)$-$\log(\Gamma_0)$ plane lies on the straight line, with gradient close to $2$.  

\begin{figure}[t!] 
\begin{center}
\scalebox{0.5}{\includegraphics{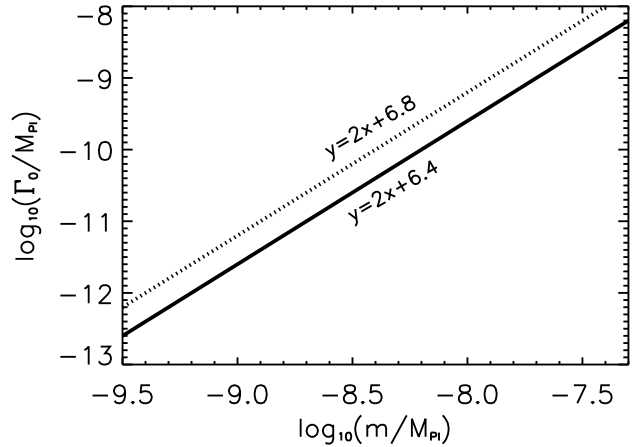}} 
\caption{\label{numthermalweak}The line of fit for the thermally weak regime, as found numerically for $n=1$ (solid line) and $n=2$ (dotted line).  
An empirical perturbation amplitude, ${\cal P_R}$, is found as given in Eqn.~(\ref{empirical}) with coefficients in Table~\ref{tab_thermalweak}.  Note that the gradient of this fit is $2$ as discussed in the text.  These lines do not change for varying values of $\phi_0$ within numerically considered ranges.}
\end{center}
\end{figure}

Numerically, we also find that, close to this line, the amplitude can be predicted using Eq.~(\ref{empirical}), but that the spectral index varies greatly.  We therefore find that the width of this line of fit (as shown in Fig.~\ref{numthermalweak}) is small, although it increases slightly with $n$, leading to a very tightly constrained fit.
\begin{table}
\begin{tabular}{|c|c|c|}
\hline
&$n=1$ &$n=2$  \\
\hline
${\cal A}$ & $5.7$ & $176.7$  \\
${\cal B}$& $3.24$ & $3.23$  \\
${\cal C}$ & $-0.19$ & $-0.69$  \\
${\cal D}$ & $-1.74$ & $-1.73$  \\
\hline
\end{tabular}
\caption{Coefficients for the thermally weak cases.  The empirical form of the amplitude is given in Eqn.~(\ref{empirical}).  Models are shown with $n=1,2$. Note that the amplitude decreases when $n$ increases due to the inverse dependence on $\Gamma$.  The shift in ${\cal C}$ is due to the weak dependence on $\phi_0$.}
\label{tab_thermalweak}
\end{table}
The best likelihood values we found are around $2\ln{\cal L}\approx 5392$, which (as we shall see) are significantly worse than $\Lambda$CDM and the remaining cases we study.  In addition, the reheat temperature in these models lies around $10^{13}$ GeV.
Due to the narrowness of the constraint, we will not consider this case in the following cosmological parameter estimation.

\subsection{Cosmological Parameter Estimation}

The fundamental observables of the dissipative inflation scenario are the same as in standard $\Lambda$CDM cosmology. These include the CMB angular power spectrum and the galaxy power spectrum. We modified the Boltzmann code CAMB \cite{Lewetal00} to calculate these observables; once the initial conditions are specified, the power spectrum observables are computed from a separate primordial power spectrum routine as described above, which bypasses the usual CAMB code without significantly increasing the computational time. Using this code, we performed cosmological parameter estimation for the four scenarios specified by weak and strong cases I and II.

We use a Markov Chain Monte Carlo (MCMC) technique \cite{Christensen:2000ji,Christensen:2001gj,Knox:2001fz,LewBri02,Kosowsky:2002zt,Verde:2003ey} to evaluate the likelihood function of model parameters. The MCMC is used to simulate observations from the posterior distribution ${\cal P}(\boldtheta|x)$, of a set of parameters $\boldtheta$ given event $x$, obtained via Bayes' Theorem,
\begin{equation}
{\cal P}(\mbox{\boldmath $\boldtheta$}|x)=
\frac{{\cal P}(x|\boldtheta){\cal P}(\boldtheta)}{\int
{\cal P}(x|\boldtheta){\cal P}(\boldtheta)d\boldtheta},
\label{eq:bayes}
\end{equation}
\noindent where ${\cal P}(x|\boldtheta)$ is the likelihood of event $x$ given the model parameters $\boldtheta$ and ${\cal P}(\boldtheta)$ is the prior probability density. The MCMC generates random draws (i.e.~simulations) from the posterior distribution that are a ``fair'' sample of the likelihood surface. From this sample, we can estimate all of the quantities of interest about the posterior distribution (mean, variance, confidence levels). A properly derived and implemented MCMC draws from the joint posterior density ${\cal P}(\boldtheta|x)$ once it has converged to the stationary distribution. We use eight chains per model and a conservative Gelman-Rubin convergence criterion \cite{gelman/rubin} to determine when the chains have converged to the stationary distribution. 

For our application, $\boldtheta$ denotes a set of cosmological parameters. We then use a modified version of the CosmoMC code \cite{LewBri02} to determine constraints placed on this parameter space by the WMAP three-year cosmic microwave background data \cite{Speetal06} and the SDSS Luminous Red Galaxy (LRG) galaxy power spectrum data \cite{Tegetal06}. In our analysis, we take the parameter set  $\{\omega_b \equiv \Omega_b h^2$, $\omega_m \equiv \Omega_m h^2$, $\theta_A, \tau, \log_{10}\left(\phi_0/\mpl\right), \log_{10}\left(\mu/\mpl\right), \log_{10}\left(\Gamma_0/\mpl\right)\}$. Here, $\theta_A$ is the angular size of the acoustic horizon and functions as a proxy for the Hubble constant $H_0\equiv 100h$ km/s/Mpc or $\Omega_{m}$. The universe is assumed to be spatially flat.  Constant priors are assumed over the previously specified parameter set, subject to the microphysical cuts applied in the dissipative inflation code as described above. We marginalize analytically over the linear bias factor $b$ and the non-linearity parameter $Q_{\rm nl}$ of the SDSS LRG data as is done normally in the CosmoMC code.

\section{Results}
\label{sec_results}

\begin{figure*}[t!]
\begin{center}
\scalebox{0.5}{\includegraphics{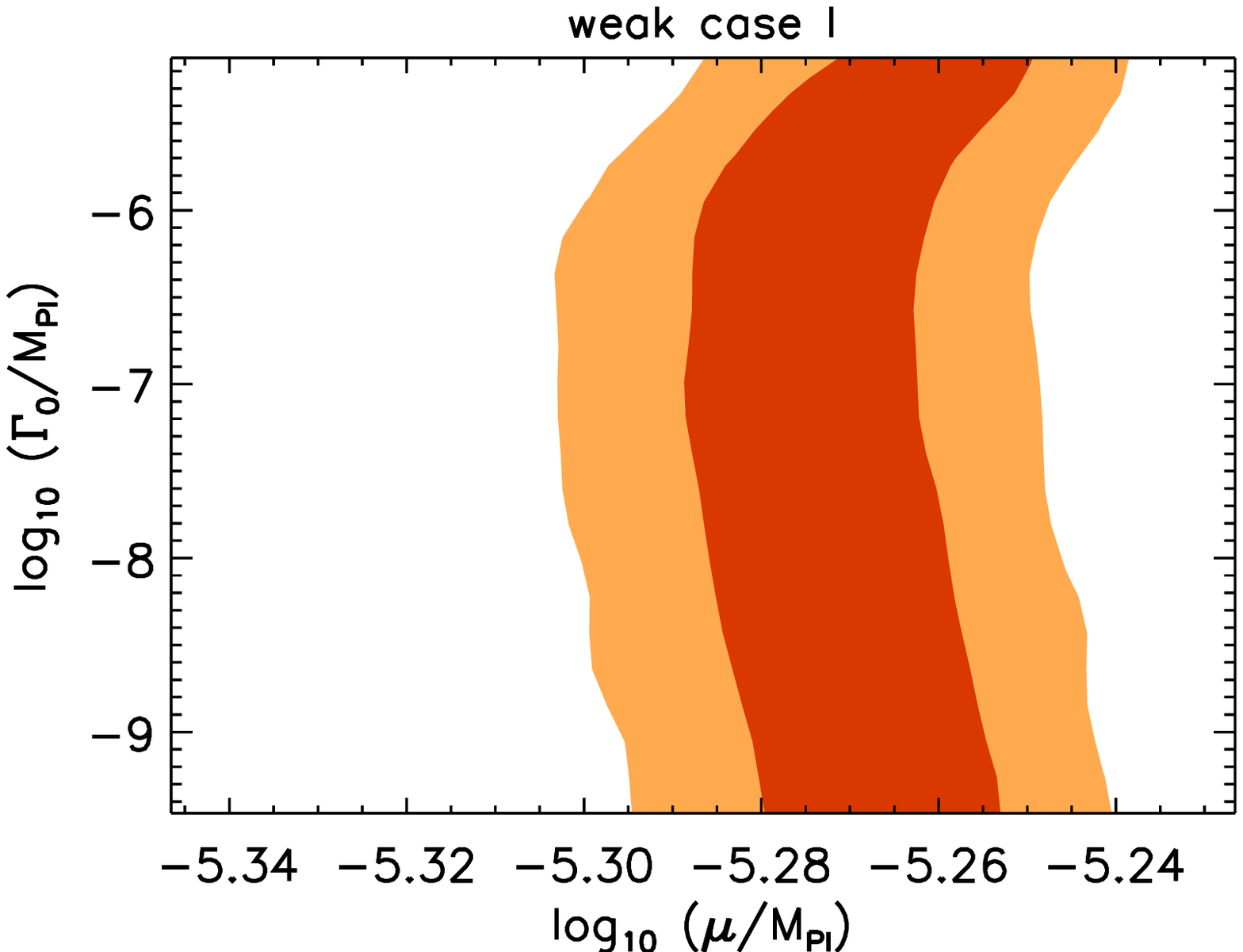}}  
\scalebox{0.5}{\includegraphics{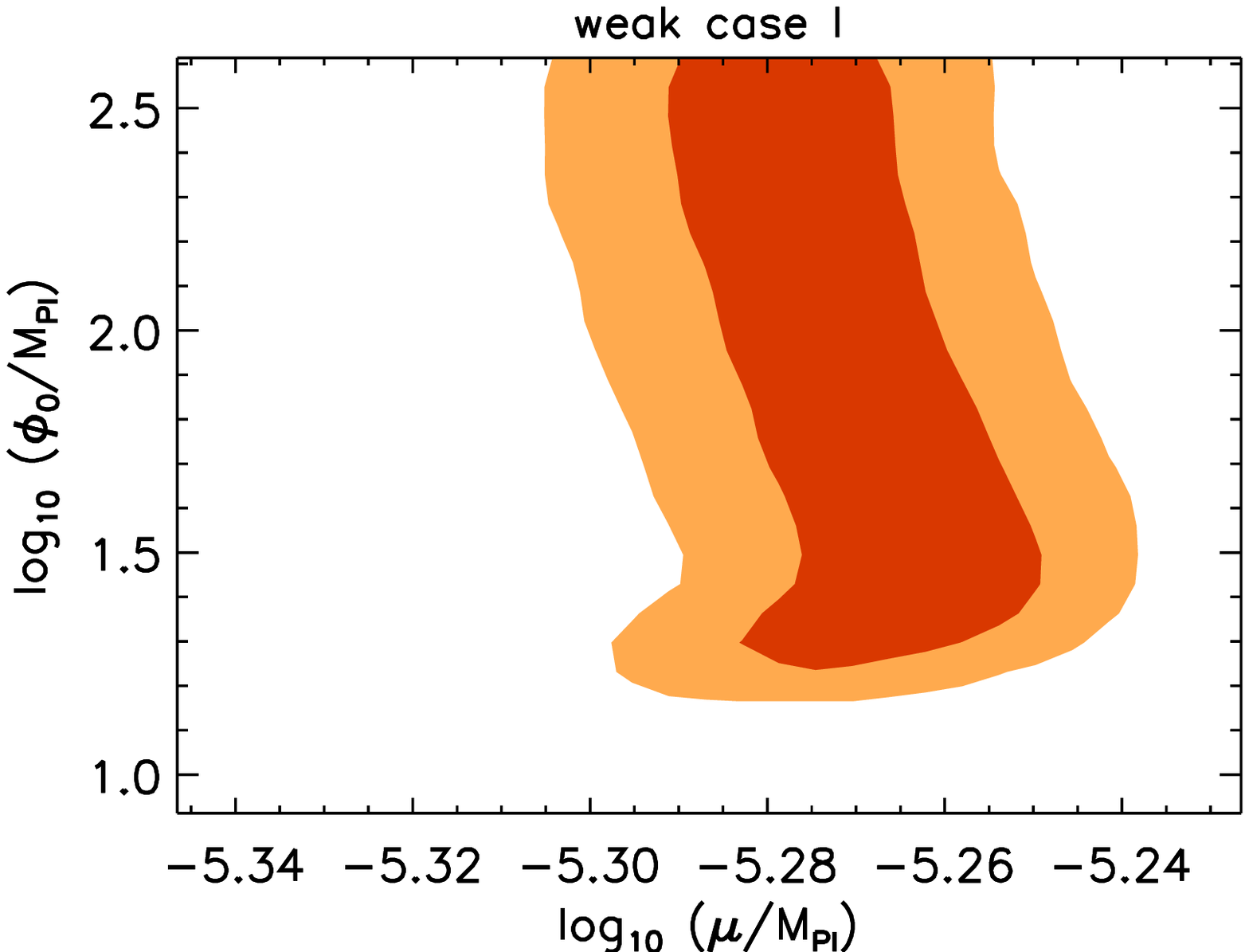}} 
\scalebox{0.5}{\includegraphics{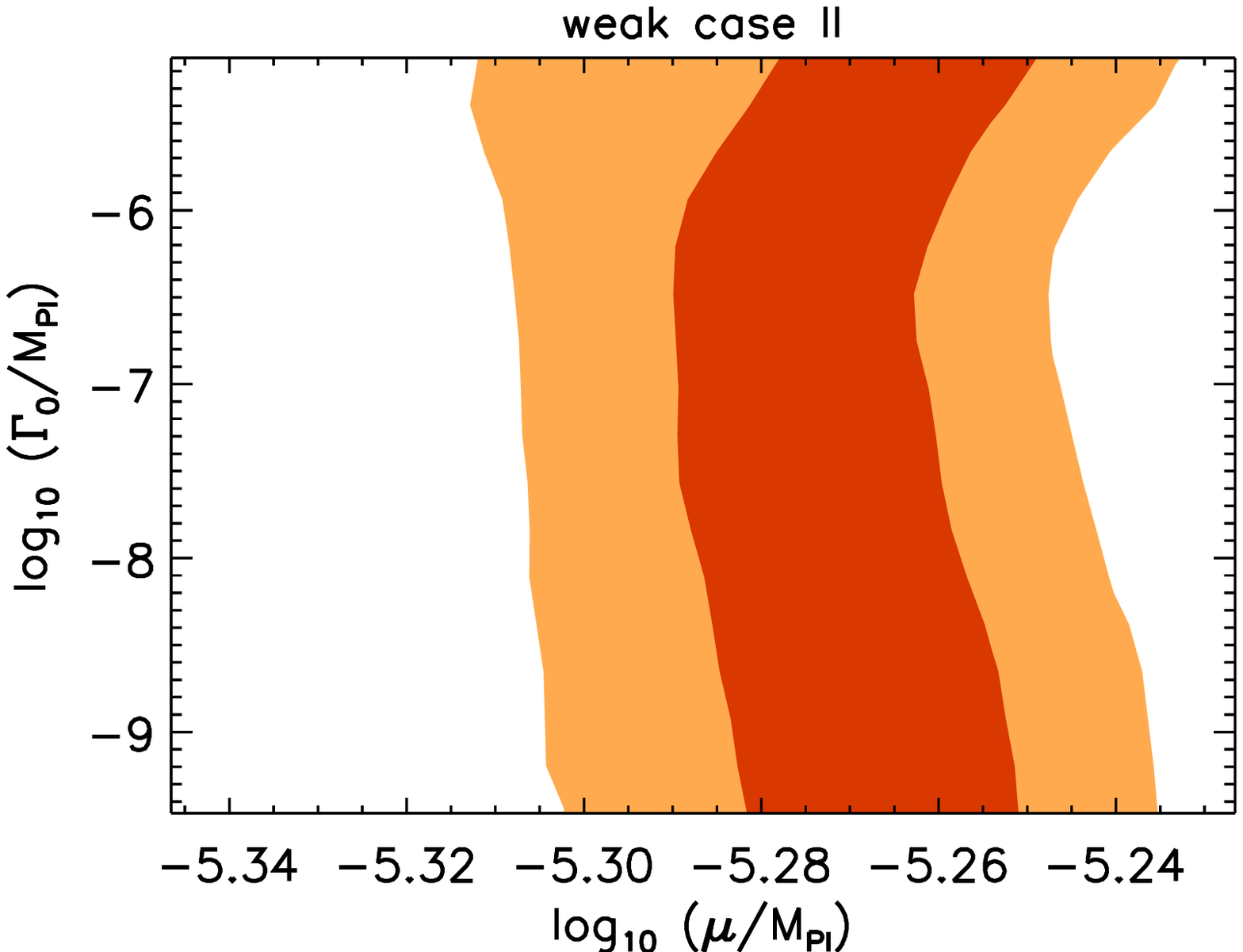}} 
\scalebox{0.5}{\includegraphics{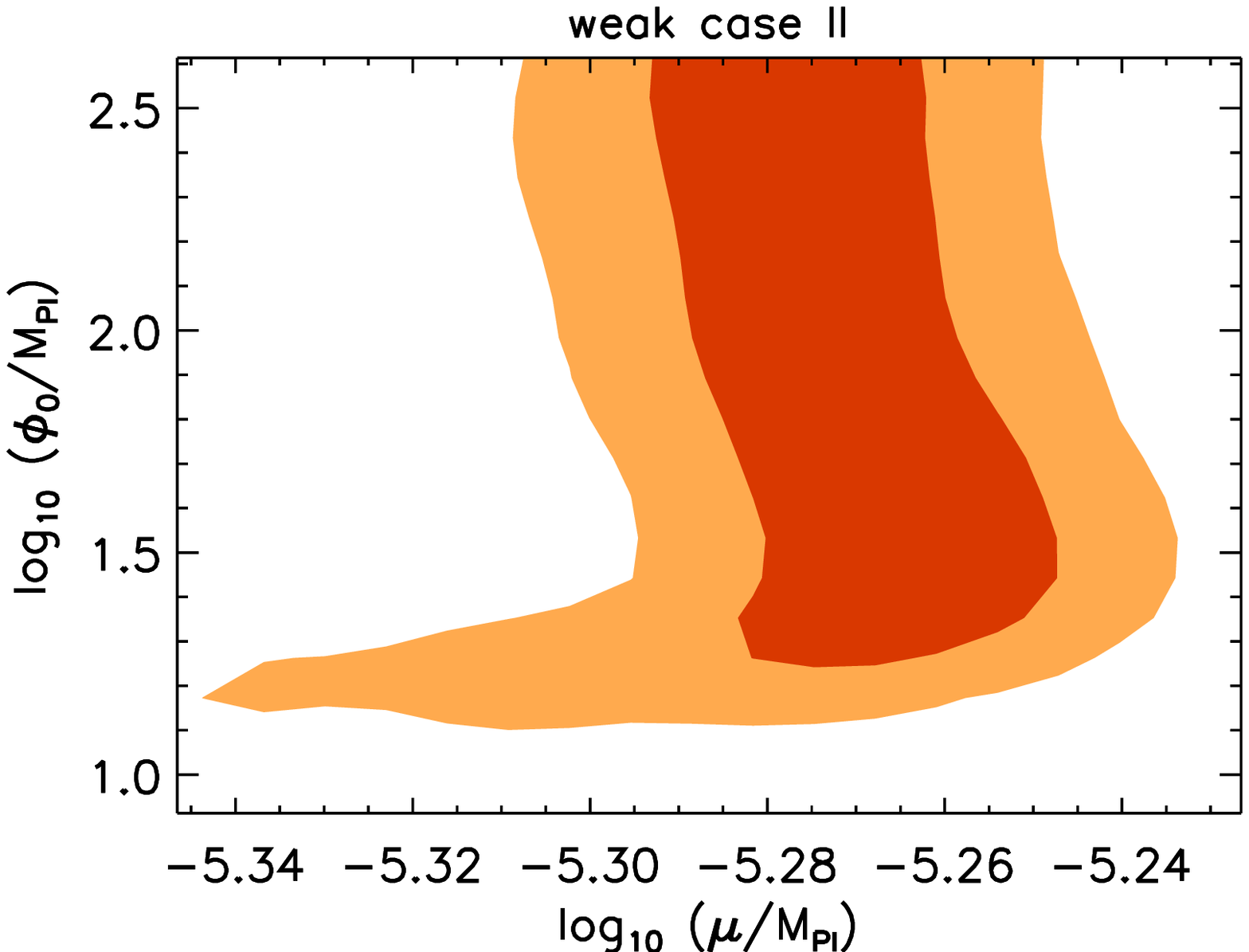}}\caption{\label{Fig3} Joint 2D 68\% (dark) and 95\% (light) CL constraints on the ``primordial'' parameters of the weak dissipation case I (left) and case II (right).}
\end{center}
\end{figure*}

\begin{figure*}[t!]
\begin{center}
\scalebox{0.5}{\includegraphics{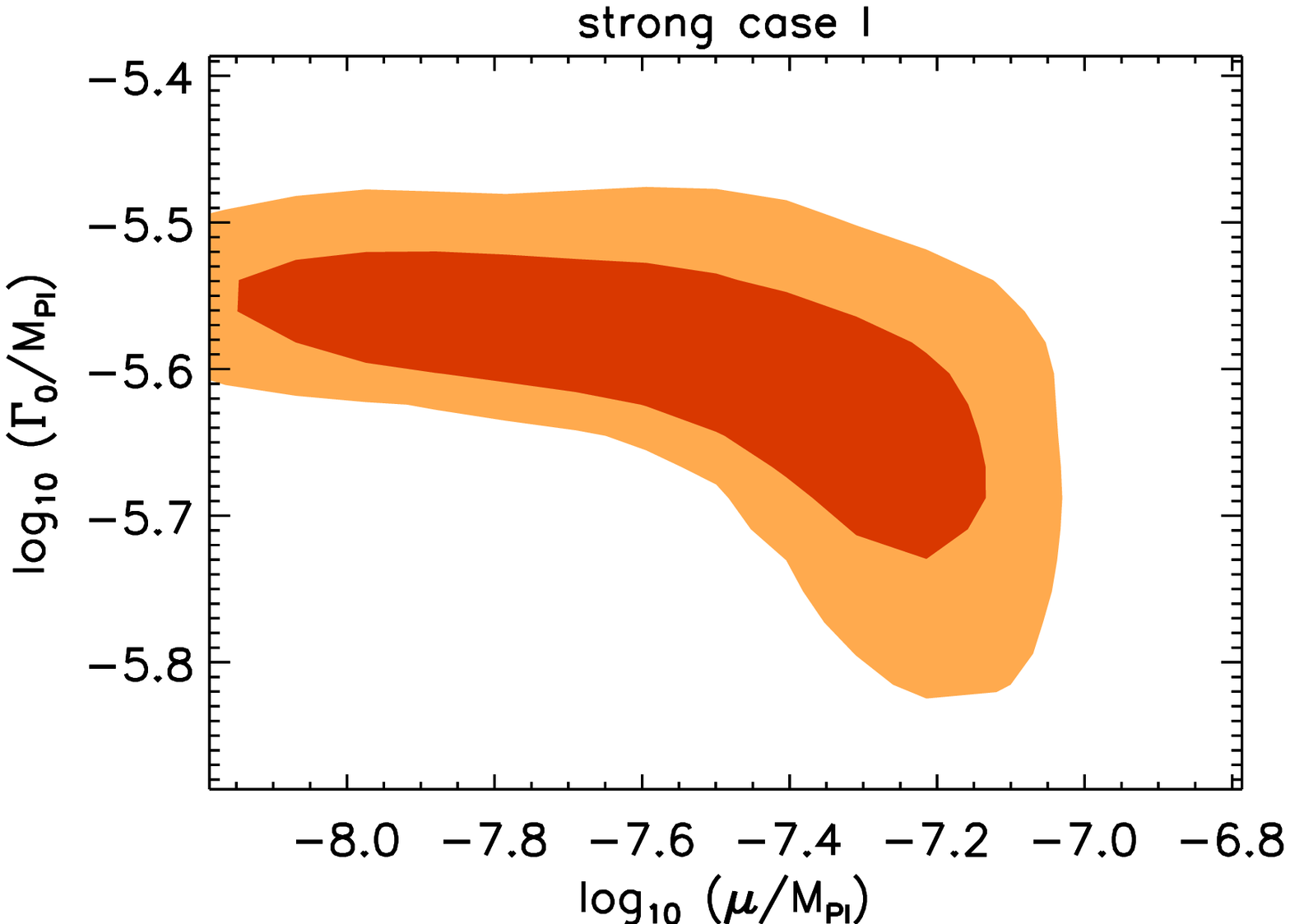}} 
\scalebox{0.5}{\includegraphics{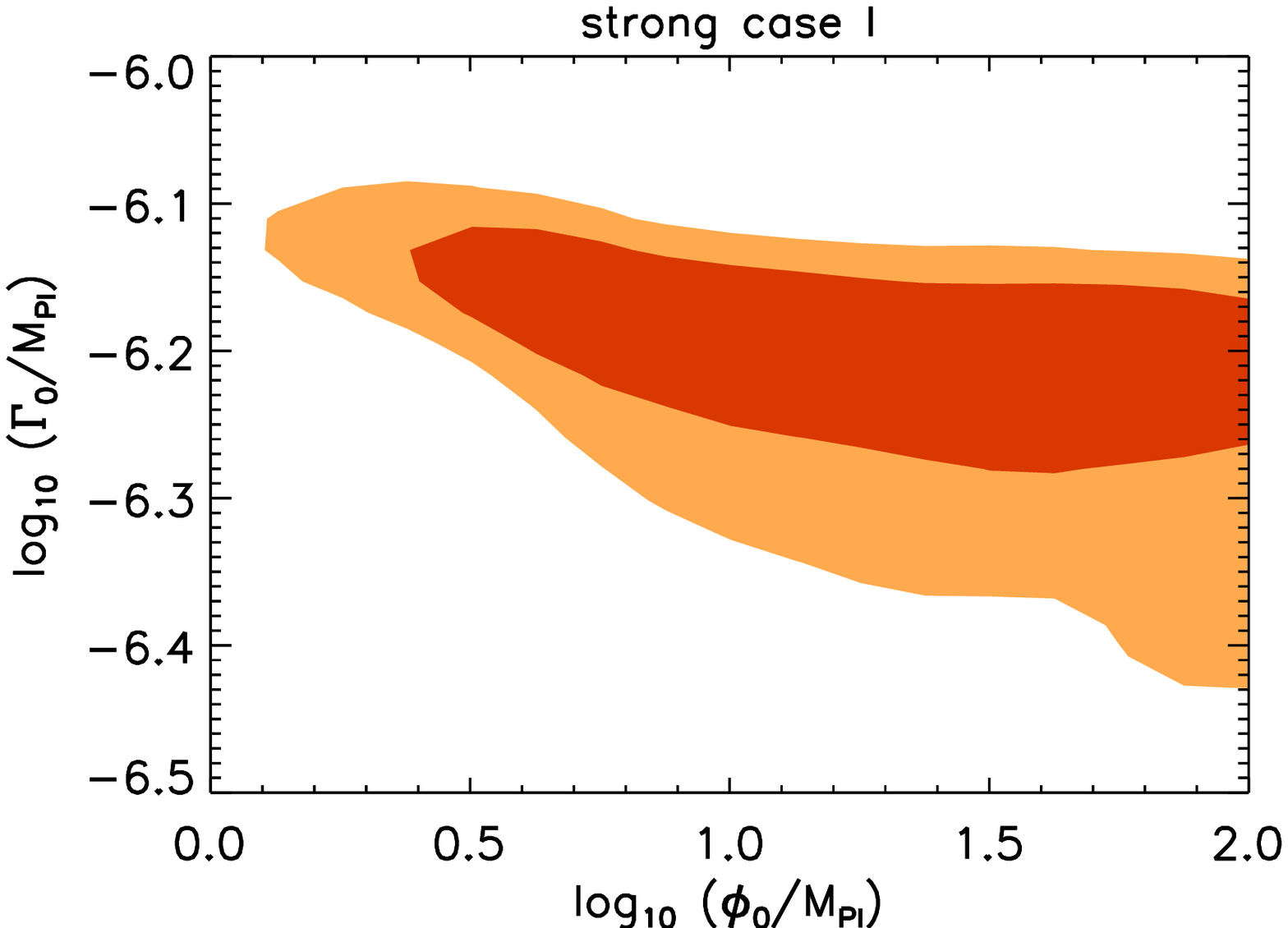}}
\scalebox{0.5}{\includegraphics{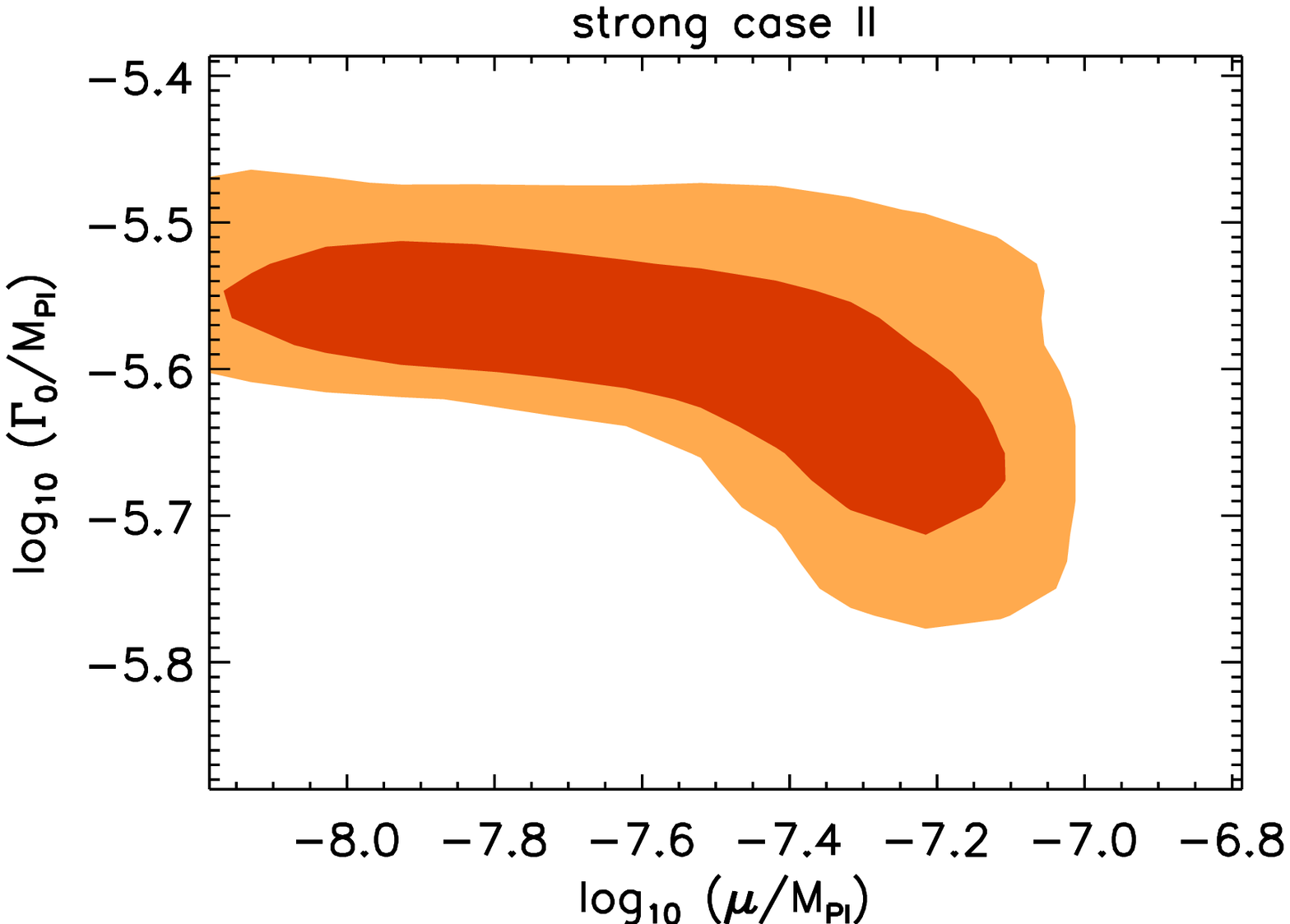}} 
\scalebox{0.5}{\includegraphics{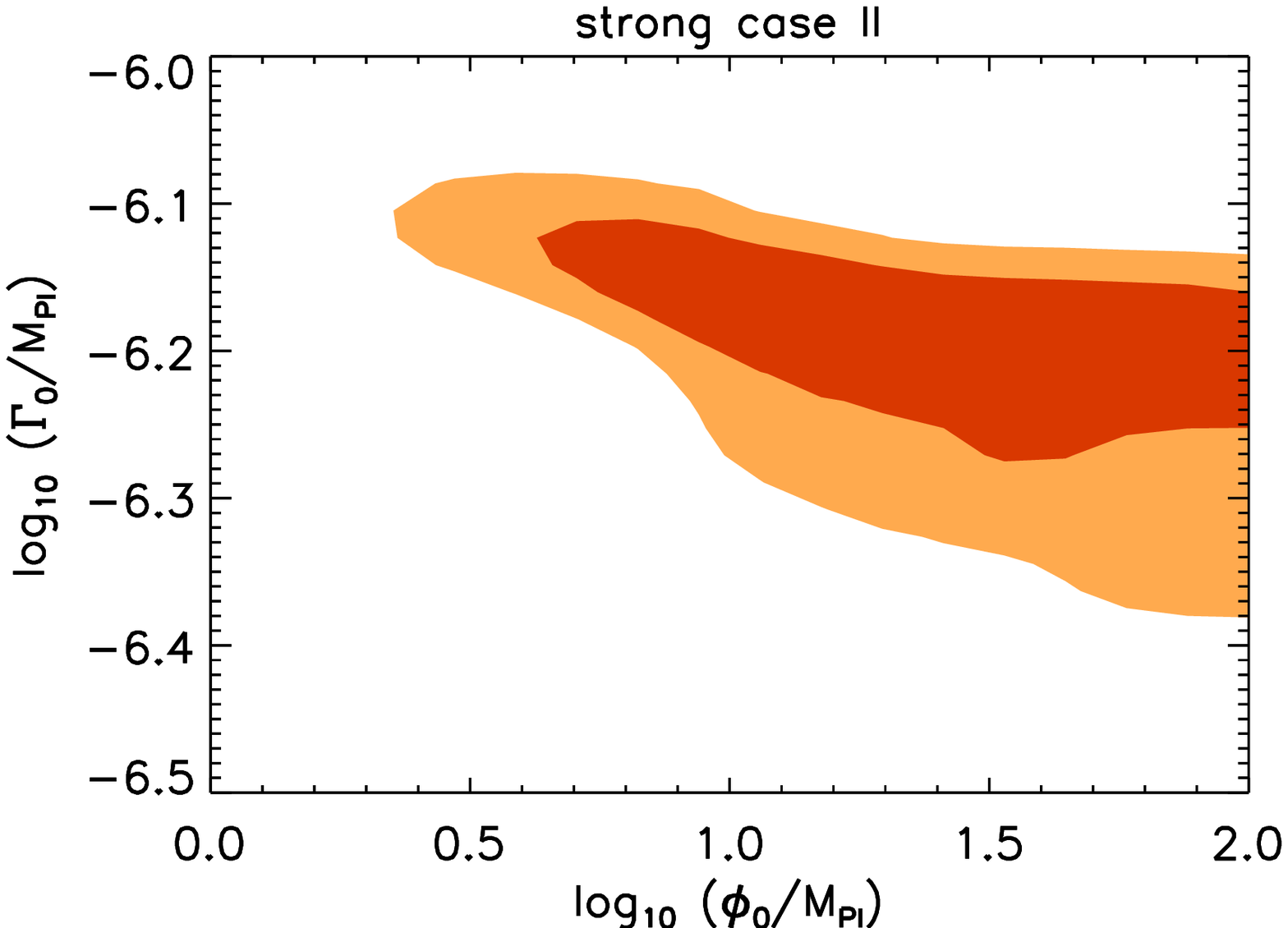}} 
\caption{\label{Fig5} Joint 2D 68\% (dark) and 95\% (light) CL constraints on the ``primordial'' parameters of the strong dissipation case I (upper) and case II (lower).}
\end{center}
\end{figure*}

\begin{figure*}[t!]
\begin{center}
\scalebox{0.5}{\includegraphics{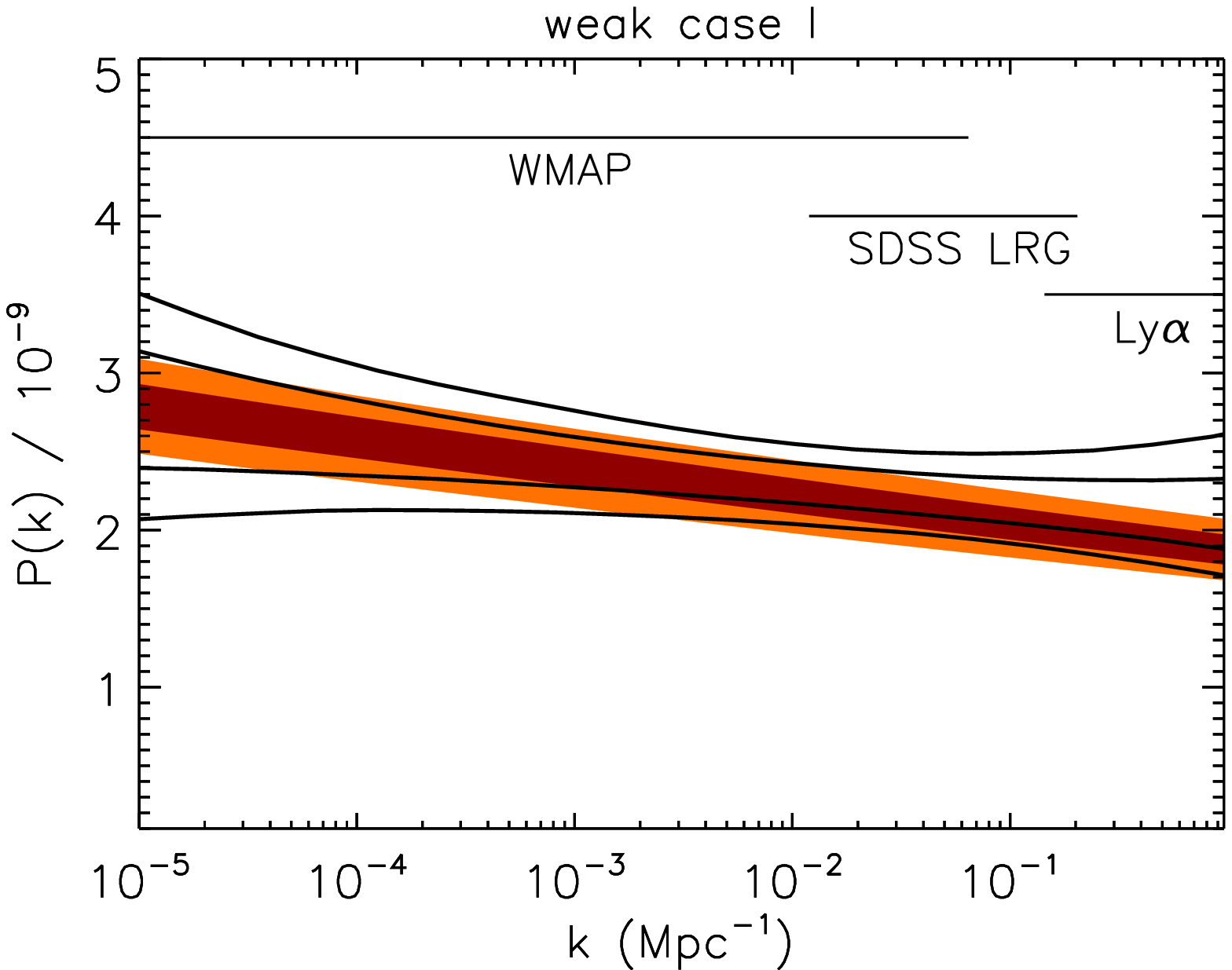}}
\scalebox{0.5}{\includegraphics{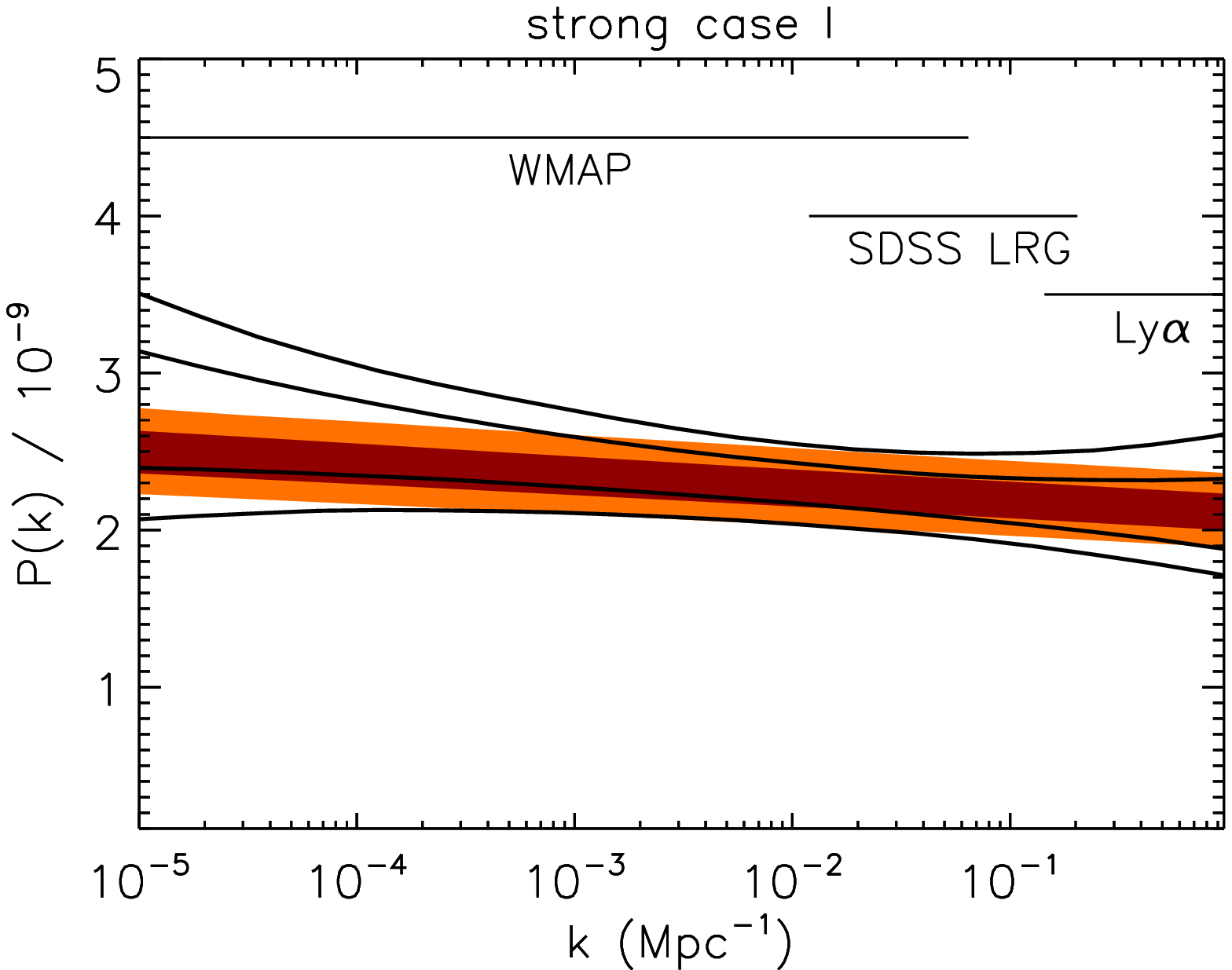}}
\caption{\label{Fig6}Reconstructed 68\% (dark) and 95\% (light) CL constraints on the primordial scalar power spectrum, shown for case I, weak (left) and strong (right). Case II constraints look very similar to their case I counterparts and are hence not shown. For comparison, the solid lines show the corresponding 68\% and 95\% constraints for single-field slow-roll inflation, taken from Ref.  \cite{Peiris:2006sj}. The range of scales spanned by WMAP and SDSS LRG data (which were used in the fit) and the smaller scales covered by Lyman--$\alpha$ data (which were not) are shown for reference. See text for discussion.}
\end{center}
\end{figure*}

In this Section, we highlight and discuss what we consider to be the most important results of our cosmological analysis. Further details can be found in Figures (\ref{numthermalweak}--\ref{Fig6}) and their captions, and in Tables \ref{Tab_chisq} and \ref{Tab_cosmo}.
It should be noted that we did not consider the thermally weak case for cosmological parameter estimation, due to the narrowness of the constraint.

Table \ref{Tab_chisq} compares the goodness of fit of the dissipative cases (weakly, thermally-weakly and strongly dissipative) studied with the ``minimal'' 6-parameter $\Lambda$CDM model. Since there is no improvement in the best fit $\chi^2$ even though the dissipative models have an extra degree of freedom, there is no indication that the data prefers the dissipative model over the $\Lambda$CDM case. 

Table \ref{Tab_cosmo} shows the 1D constraints on the ``late-time'' cosmological parameters for the weak and strong cases after all other parameters have been marginalized over. The marginalized probability distribution functions for these parameters are very close to Gaussian, and they do not shift from the standard $\Lambda$CDM concordance values; they are all minimally correlated with the ``early-time'' dissipative parameters.
We also expect this to generalise to the thermally weak scenario.

Figures (\ref{Fig3}) and (\ref{Fig5}) show the joint 2D constraints on the ``primordial'' dissipative parameters for the weak and strong cases respectively. Unlike the ``late-time'' cosmological parameters, these parameters have highly non-Gaussian likelihood surfaces and are also very correlated. 

As predicted by the analytics in Section \ref{SecWeak}, for the weak cases, only $\mu$ is constrained; both $\Gamma_0$ and $\phi_0$ extend past their numeric priors and there is no correlation between these parameters. For both cases I and II, $\log_{10}\left(\mu/\mpl\right)\approx -5.27$, close to the analytic prediction.  Additionally, the results are almost exactly independent of $n$ (i.e. case independent) as expected.
The sharp bend in the contour (for large $\Gamma_0$) occurs due to the increase in the friction ratio, $r$, at these values (seen on the LHS of Figure~(\ref{Fig2})), as shown by Eqn. (\ref{analyticm}), and is hence slightly dependent on the form of $\Gamma$ in this region.

There is also a noticeable shift in the mass, $m$, when $\phi_0\sim\mpl$.  The low $\phi_{0}$ values occur when the total number of $e$-foldings, $N$, is just large enough (i.e. $N\sim60$), as seen by Eqn.~(\ref{efoldings}).  
For these values, horizon crossing occurs very close to the top of the potential.  
Additionally, when $\phi_0\rightarrow\mpl$ from above, the slow-roll factor, $\epsilon$, increases and therefore the spectral index decreases.  In order to match observation, this shift must be therefore balanced by a shift in $m$.
The case in which cosmological scales cross the horizon at the top of the potential is unphysical and is not of relevance.

For strong dissipation, as predicted in Section \ref{SecStrong}, only $\Gamma_0$ is tightly constrained. When $r\gg1$, $\phi_*\approx\phi_0$, and $\log_{10}\left(\Gamma_0/\mpl\right)\approx -5.55$ as given by Eqn. (\ref{analyticG}) with almost no $\phi_0$ dependence.  
This may be compared with the numerical results of \cite{Hall:2004ab}, in which a reasonable fit was found for $[m/\mpl,\phi_0/\mpl,\Gamma_0/\mpl]\approx[10^{-9},0.8,10^{-6}]$ (compared to WMAP I data).
As $r\rightarrow1$ ($\mu$ increasing), the correlation between $\Gamma_0$ and $\phi_0$ is non-negligible and can be seen numerically.
There is an upper constraint on $\mu$ due to the assumption of $r\geq 1$, but the lower limit should extend {\it ad infinitum}; the lower bound is derived from the $N<10^5$ bound we imposed. 
The upper bound of $\phi_0/\mpl<2$ is also due to this $e$-folding constraint, as predicted.

It should be noted that for both weak and strong cases, there is a lower bound of $\log_{10}\left(\phi_0/\mpl\right)>0 $, showing that the data disfavours $\phi_0<\mpl$. The constraint is slightly stronger for the weak case.  This is due to the increase in the slow-roll parameter, $\epsilon$, which leads to an increase in the tilt (the index becomes redder).  
As this index become redder than the data, the fit becomes worse.

Figure (\ref{Fig6}) shows the primordial scalar power spectra $P(k)$ reconstructed from the MCMC runs using WMAP 3 year and SDSS LRG data for dissipative inflation. For comparison, the reconstructed scalar power spectrum results of Figure (10) of Ref. \cite{Peiris:2006sj} (see bottom right panel) are shown. The latter reconstruction was done using WMAP 3 year data and the SDSS main galaxy sample power spectrum, under the assumption that the primordial fluctuations are seeded by the standard single-field slow roll inflation mechanism that additionally satisfied a minimal ``sufficient $e$-folds'' requirement that solves the cosmological flatness and horizon problems. The dissipative constraints are tighter than the single-field slow roll constraints for two reasons: (a) the SDSS LRG sample is more constraining than the SDSS main galaxy sample, and (b) the single-field slow roll analysis marginalizes over the shape of the inflationary potential using a Hamilton-Jacobi formalism \cite{Peiris:2006ug}. 

Similar to the single-field slow roll constraints of  \cite{Peiris:2006sj}, and consistent with the empirical power-law fits of \cite{Speetal06} and \cite{Tegetal06}, the dissipative constraints on $P(k)$ show a distinct preference for a \emph{red} power spectrum. However, there is a systematic difference in the reconstructed shape compared to the single-field slow roll case. 
The red spectra are predicted by the analytics in Eqn. (\ref{indices}) for both weak and strong cases.  Due to the $r$-suppression in the strong cases, the tilt is expected to be less red than for the weak cases and this is what we see in the reconstruction. 
No significant running is seen as expected.

\begin{table}
\begin{center}
\begin{tabular}{|c|c|}
\hline 
Model & Best fit $-2 \ln\cal{L}_\mathrm{max}$ (WMAP+SDSS LRG) \\ 
\hline \hline
$\Lambda$CDM   &             5374.58 \\
 \hline
Weak I    &    5374.78\\
 \hline
Weak II    &    5374.75 \\
\hline
Thermally Weak  &      $\sim$5392 \\
 \hline
Strong I  &      5375.19 \\
 \hline
Strong II  &      5376.45 \\
\hline
\end{tabular}
\end{center}
\caption{The best fit chi square, defined as $\chi^2 = -2 \ln\cal{L}_\mathrm{max}$ (where $\cal{L}_\mathrm{max}$ is the maximum likelihood with respect to the WMAP 3 year data and the SDSS LRG galaxy power spectrum data) for the standard $\Lambda$CDM model and the four dissipative inflation scenarios analyzed in this work. The $\Lambda$CDM model gives a slightly better (lower) $\chi^2$ for this dataset than any of the dissipative inflation scenarios considered, while containing one less degree of freedom, and hence the data does not exhibit a preference for any of the dissipative models.}
\label{Tab_chisq}
\end{table}

\begin{table*}
\begin{center}
\begin{tabular}{|c|c|c|c|c|}
\hline 
Parameter & Weak I & Weak II & Strong I & Strong II  \\ 
\hline \hline 
$\Omega_b h^2$ & $0.02234^{+0.00043}_{-0.00044}$ &  $0.02234^{+0.00045}_{-0.00046}$ &  $0.02302 \pm 0.00047$ & $0.02304^{+0.00048}_{-0.00047}$\\
\hline
$\Omega_c h^2$ &  $0.1066 \pm 0.0042$ &  $0.1066 \pm 0.0043$ & $0.1070 \pm 0.0044$ & $0.1070 \pm 0.0044$  \\
\hline
$\tau$ &  $0.092 \pm 0.026$ &  $0.091 \pm 0.026$ & $0.109 \pm 0.025$ &  $0.110^{+0.027}_{-0.028}$  \\
\hline
$h$ & $0.729^{+0.015}_{-0.016}$ &  $0.729^{+0.015}_{-0.016}$ & $0.726^{+0.017}_{-0.016}$ &  $0.742^{+0.016}_{-0.017}$ \\
\hline
\end{tabular}
\end{center}
\caption{Constraints on the ``late-time'' cosmological parameters from the WMAP and SDSS LRG data-sets (mean, upper and lower 68\% CL, marginalizing over all other parameters), for the four dissipative inflationary scenarios described in the text.}
\label{Tab_cosmo}
\end{table*}

Numerically we have checked that the reheat temperatures from the simulated models and the analytical approximation agreed well.  
For example, the strong chains have reheat temperatures around $T_R\approx10^{12}$-$10^{14}$GeV.

Since the numerical code does not simulate supercooled inflation, the constraints in this paper do not display a supercooled limit.  It should be noted, however, that it is impossible with this potential to obtain enough supercooled inflation to solve the standard problems (horizon, flatness);  the parameters never reach a "supercooled limit" in the MCMC.

\section{Conclusion}
We have studied the dissipative inflationary regime when the dissipative term takes a specific form, $\Gamma=\Gamma(\phi)$.  Two forms of $\Gamma$ have been analyzed in the weak, thermally weak and strong dissipative regimes, assuming a SUSY breaking form for the potential. Our system introduces three new parameters; two for the potential and one for the dissipative term.

We derive analytic formulae which relate
the mass scale and scale of dissipation to the characteristic amplitude of the primordial power spectrum and its shape.
In the weak regime, the level of dissipation is unconstrained but the mass scale is set around $10^{-6}\mpl$ (using typical WMAP values).
Conversely, in the strong regime, the mass scale is unconstrained, but the dissipative coefficient is tightly constrained around $10^{-6}\mpl$.
In the thermally weak regime, a non-trivial relation between all three parameters is found for the freeze-out value of the perturbation amplitude.  Oscillation in $\delta\phi$ across the horizon, in this latter case, means that the final amplitude cannot be analytically approximated and an empirical fit has been found.

Using a numerical simulation of dissipative inflation to calculate the observable primordial power spectrum exactly, we explore constraints on the dissipative parameter space given by the WMAP 3 year data and the SDSS LRG power spectrum using Markov Chain Monte Carlo techniques. We find that the values of the cosmological parameters do not shift significantly from the standard $\Lambda$CDM concordance values. The constraints on the dissipative parameters show large correlations and agree well with expectations from the analytic understanding; the mass scale alone is constrained for weak dissipation and only the strength of dissipation is constrained in the strong dissipative regime. In both cases, a sub-Planckian inflaton is disfavoured by the data. In both weak and strong regimes, we reconstruct the primordial power spectrum and show that these models prefer a {\it red} spectrum, with no obvious running of the spectral index. Despite the addition of an extra degree of freedom compared to the minimal concordance cosmology, the data does not display a preference for any of the dissipative models; the best fit is comparable to the standard $\Lambda$CDM model. 

In the models considered in this paper, the reheat temperature is between $T_R\approx10^{10}$--$10^{14}$~GeV, which does not satisfy present constraints for gravitino production.  This temperature can be reduced by raising the ratio of dissipation to the Hubble parameter.  We also comment on the level of non-Gaussianity predicted in these models and relate them to our dissipative parametrization.

This analysis does not consider a thermal dissipative term $\left(\Gamma\ne\Gamma(T)\right)$, which is expected to lead to oscillations in the primordial power spectrum.  These terms are beyond the scope of this work and will be the focus of future work.

\smallskip
{\it Acknowledgments:} The authors wish to thank Ma Bastero-Gil, Arjun Berera, Anthony Brookfield and Ian Moss for many useful discussions, and Antony Lewis for discussions on efficient samplers for highly non-Gaussian distributions.
We would also like to thank Carsten van de Bruck and Richard Easther for comments on the draft.
We are grateful for the hospitality of the Institute of Astronomy, Cambridge, UK where part of this work was carried out.  We acknowledge use of the Legacy Archive for Microwave Background Data Analysis (LAMBDA). This work was partially supported by the National Center for Supercomputing Applications under grant number TG-AST070004N and utilized computational resources on the TeraGrid (Cobalt). LH is supported by a PPARC Postdoctoral Fellowship. HVP is supported by NASA through Hubble Fellowship grant \#HF-01177.01-A from the Space Telescope Science Institute, which is operated by the Association of Universities for Research in Astronomy, Inc., for NASA, under contract NAS 5-26555.

\bibliography{paper.bib}

\end{document}